\documentclass[12pt, draftclsnofoot, onecolumn]{IEEEtran}
\usepackage[T1]{fontenc}

\usepackage{graphicx}
\usepackage[noadjust]{cite}
\usepackage{mcite}
\usepackage{amsfonts,helvet}
\usepackage{fancyhdr}
\usepackage{threeparttable}
\usepackage{epsf,epsfig}
\usepackage{amsthm}
\usepackage{amsmath}
\usepackage{siunitx}
\usepackage{amssymb}
\usepackage{dsfont}
\usepackage{subfigure}
\usepackage{color}
\usepackage{enumerate}
\usepackage{hyperref}
\usepackage{cancel}
\usepackage{bbm}
\usepackage{dsfont}
\usepackage[subnum]{cases}
\usepackage{adjustbox}
\usepackage[linesnumbered,ruled,noend]{algorithm2e}
\usepackage{multicol}
\usepackage[english]{babel}
\usepackage{enumitem}

\newtheorem{theorem}{Theorem}

\newtheorem{corollary}{Corollary}

\newmuskip\pFqmuskip

\def\bb{{\bf b}}

\def\bff{{\bf f}}
\def\bg{{\bf g}}
\def\bh{{\bf h}}

\def\bn{{\bf n}}

\def\bq{{\bf q}}
\def\br{{\bf r}}
\def\bs{{\bf s}}

\def\bu{{\bf u}}

\def\bw{{\bf w}}
\def\bx{{\bf x}}
\def\by{{\bf y}}

\def\bA{{\bf A}}

\def\bC{{\bf C}}
\def\bD{{\bf D}}
\def\bE{{\bf E}}
\def\bF{{\bf F}}
\def\bG{{\bf G}}
\def\bH{{\bf H}}
\def\bI{{\bf I}}

\def\bM{{\bf M}}

\def\bW{{\bf W}}

\def\cC{\mbox{$\mathcal{C}$}}

\def\cL{\mbox{$\mathcal{L}$}}

\def\cN{\mbox{$\mathcal{N}$}}

\def\cP{\mbox{$\mathcal{P}$}}
\def\cQ{\mbox{$\mathcal{Q}$}}

\def\bbC{\mbox{$\mathbb{C}$}}

\def\bbR{\mbox{$\mathbb{R}$}}

\def\ubG{\mbox{$\underline{\bf{G}}$}}
\def\ubH{\mbox{$\underline{\bf{H}}$}}

\def\ubM{\mbox{$\underline{\bf{M}}$}}

\def\ubW{\mbox{$\underline{\bf{W}}$}}

\def\ubg{\mbox{$\underline{\bf{g}}$}}

\def\ubn{\mbox{$\underline{\bf{n}}$}}

\def\ubq{\mbox{$\underline{\bf{q}}$}}

\def\ubs{\mbox{$\underline{\bf{s}}$}}

\def\ubu{\mbox{$\underline{\bf{u}}$}}

\def\ubw{\mbox{$\underline{\bf{w}}$}}
\def\ubx{\mbox{$\underline{\bf{x}}$}}
\def\uby{\mbox{$\underline{\bf{y}}$}}

\def\btau{\boldsymbol{\tau}}
\def\bSigma{\boldsymbol{\Sigma}}

\usepackage{array}

\makeatletter
\newcommand{\thickhline}{%
    \noalign {\ifnum 0=`}\fi \hrule height 1pt
    \futurelet \reserved@a \@xhline
}
\newcolumntype{"}{@{\hskip\tabcolsep\vrule width 1pt\hskip\tabcolsep}}
\makeatother

\IEEEoverridecommandlockouts

\title{\huge
Quantized Massive MIMO Systems with Multicell Coordinated Beamforming and Power Control
}

\author{
Jinseok Choi, {\it Member, IEEE,} Yunseong Cho, {\it Student Member, IEEE,} \\
and Brian L. Evans, {\it Fellow, IEEE} \thanks{
J. Choi is with Qualcomm Inc. Wireless R\&D, San Diego, CA 92121 USA
(e-mail:jinseokchoi89@utexas.edu).
Y. Cho and B. L. Evans are with the Wireless Networking and Communication Group (WNCG), Department of Electrical and Computer Engineering, The University of Texas at Austin, Austin, TX 78701 USA. (e-mail: 
yscho@utexas.edu, bevans@ece.utexas.edu).
}
}

\begin{document}
\maketitle

\begin{abstract}

In this paper, we investigate a coordinated multipoint (CoMP) beamforming and power control problem for base stations (BSs) with a massive number of antenna arrays under coarse quantization at low-resolution analog-to-digital converters (ADCs) and digital-to-analog converter (DACs). 
%
%
Unlike high-resolution ADC and DAC systems,  non-negligible quantization noise that needs to be considered in CoMP design makes the problem more challenging.
%
We first formulate total power minimization problems of both uplink (UL) and downlink (DL) systems subject to signal-to-interference-and-noise ratio (SINR) constraints.
We then derive strong duality for the UL and DL problems under coarse quantization systems.
%
Leveraging the duality, we propose a framework that is directed toward a twofold aim: to discover the optimal transmit powers in UL by developing iterative algorithm in a distributed manner and to obtain the optimal precoder in DL as a scaled instance of UL combiner.
%
Under homogeneous transmit power and SINR constraints per cell, we further derive a deterministic solution for the UL CoMP problem by analyzing the lower bound of the SINR.
Lastly, we extend the derived result to wideband orthogonal frequency-division multiplexing systems to optimize transmit power and beamformer for all subcarriers.
Simulation results validate the theoretical results and proposed algorithms.

\end{abstract}
\begin{IEEEkeywords}
Coordinated multipoint, joint beamforming and power control,  low-resolution ADC/DAC, total transmit power minimization, UL-DL strong duality
\end{IEEEkeywords}

\section{Introduction}
\label{sec:intro}

Employing large-scale antenna arrays at the BS has been widely studied in last decades as a potential future wireless communication technology because of its significant gain in spectral efficiency \cite{marzetta2010noncooperative}.
Due to the large number of antennas followed by power-demanding high-resolution analog-to-digital converters (ADCs) and digital-to-analog converters (DACs), however, significant power consumption becomes one of the primary practical challenges in realizing the system.
Accordingly, employing low-resolution quantizers has attracted the most interest as a low-power solution in recent years \cite{jacobsson2017quantized,li2020interference, choi2016near, studer2016quantized}.
In multicell systems, non-negligible quantization error due to the low-resolution quantizers is a function of not only the in-cell channels and beamformers but also the inter-cell channels and beamformers. 
In this regard, we investigate coordinated multipoint (CoMP) beamforming (BF) and power control (PC) problems in low-resolution massive multiple-input and multiple-output (MIMO) systems to take into account the effect of the quantization error to the beamformer design and power allocation in the multicell communications.

\subsection{Prior Work}

As modern cellular systems operate on the interference-limited regime, the coordination between base stations (BSs) has shown large gain in communication performance \cite{rashid1998joint,rashid1998transmit,bengtsson1999optimal,stridh2006system,ng2008distributed,song2007network,dahrouj2010coordinated,sanderovich2007uplink,irmer2011coordinated,shin2017coordinated,jungnickel2014role}.
Problems of minimizing transmit power for given quality of service constraints were often investigated in multicell CoMP networks.
In \cite{rashid1998joint}, an uplink (UL) BF and PC method was developed by utilizing a fixed point iteration method.
In addition, a downlink (DL) BF and PC method was further proposed in \cite{rashid1998transmit}. 
Due to the difficulty in designing DL BF, the DL BF was derived by exploiting a virtual UL concept based on the duality between UL single-input and multiple-output and DL multiple-input and single-output systems.
In \cite{bengtsson1999optimal}, relaxing and casting the DL BF problem into a semidefinite optimization problem, a DL BF solution was efficiently computed by using interior point methods.
In addition to the semidefinite relaxation optimization for the BF design, BS allocation and congestion control were further investigated in \cite{stridh2006system}, providing substantial increase in the system performance.
Assuming interference only from adjacent cells, a Kalman smoothing based BF method was developed by recasting the DL BF problem to a virtual minimum mean-squared error (MMSE) estimation problem to design network-wide MMSE BF without requiring a central processing unit \cite{ng2008distributed}.
Linear programming-based network duality for MIMO UL and DL with a single layer was leveraged in \cite{song2007network} to develop more efficient BF algorithms both in convergence and performance.
Lagrangian based duality for multiuser MIMO systems was further derived in \cite{dahrouj2010coordinated} and used to propose an distributed algorithm, requiring less synchronization and complexity burden on users and BSs.
Practical contraints such as limited backhaul capacity was considered in \cite{sanderovich2007uplink}, and a CoMP BF system was implemented in a real field testbed in \cite{irmer2011coordinated}, showing its benefits in spectral efficiency especially for cell edge users. 
Improving the data rates of cell-edge users, a CoMP BF problem based on interference alignment was also studied in a non-orthogonal multiple access system \cite{shin2017coordinated}. 
Recently, understanding the benefit of employing a large antenna arrays at the BS, the performance gain from using massive antenna arrays jointly with CoMP BF was demonstrated by providing a more robust link and more localized interference \cite{jungnickel2014role}.
Although prior findings in MIMO CoMP systems can be naturally extended to massive MIMO systems with high-resolution ADCs and DACs, employing low-resolution ADCs and DACs further needs to be considered to address the excessive power consumption problem.

To achieve power-efficient communications, low-resolution ADC architectures have been extensively investigated in recent years \cite{choi2016near, studer2016quantized,wang2017bayesian,wang2019reliable,jeon2018supervised,choi2019robust,li2017channel,jacobsson2017throughput, zhang2016mixed, choi2017resolution, orhan2015low, xu2019uplink, choi2019two}.
As an effort to realize low-resolution ADC systems, essential wireless communication techniques such as channel estimation and detection have been developed in low-resolution ADC systems \cite{choi2016near, studer2016quantized,wang2017bayesian,wang2019reliable,jeon2018supervised, choi2019robust,li2017channel, jacobsson2017throughput}.
Unified frameworks for channel estimation and symbol detection were developed for 1-bit ADC systems by using 1-bit maximum likelihood estimation \cite{choi2016near}.
Quantized maximum a-posteriori channel
estimation and data detection were also developed by showing that 4-bit ADCs yield no performance loss from infinite-resolution ADCs \cite{studer2016quantized}.
For orthogonal frequency-division multiplexing (OFDM) systems, a generalized turbo estimator was utilized for channel estimation and symbol detection with  over-the-air experiments, showing reasonable reliability when using low-resolution ADCs \cite{wang2017bayesian,wang2019reliable}.
In \cite{jeon2018supervised, choi2019robust}, learning-based detectors were proposed without requiring explicit channel estimation.
As a special low-resolution ADC system, a detector for mixed-ADC systems was proposed in \cite{zhang2016mixed}.
In addition, a resolution-adaptive ADC system was proposed with near optimal bit-allocation solutions \cite{choi2017resolution}.
For tractability, linear quantization models such as Bussgang decomposition \cite{li2017channel, jacobsson2017throughput} and an additive quantization noise model (AQNM) \cite{orhan2015low, xu2019uplink, choi2019two}  were utilized by providing insightful analytical results.

Low-resolution DAC systems have also been studied in many literatures \cite{jacobsson2017quantized,wang2018finite,li2020interference}.
Achievable rates with linear precoders were derived in low-resolution DAC systems, and a nonlinear precoder was developed for 1-bit DAC systems, showing that using 3-4 bits offers comparable performance to infinite-resolution DACs and that the proposed 1-bit precoder causes only 3 dB loss from infinite-resolution DACs \cite{jacobsson2017quantized}.
A universal precoding approach was further developed in \cite{wang2018finite} by improving the performance and complexity trade-off from \cite{jacobsson2017quantized}.
The rate analysis in \cite{li2017downlink} showed that using $ 2.5\times$ more antennas can compensate for performance loss due to using 1-bit DACs. 
In addition, a constructive interference approach was adopted in \cite{li2020interference} to develop a low-complexity precoder for 1-bit DAC systems.
For orthogonal frequency-division multiplexing (OFDM) systems, the rate and bit-error-rate (BER) analysis in \cite{jacobsson2019linear} demonstrated that using 3-4 bits can achieve the BER  comparable to that of infinite-resolution DAC systems.
A mixed-DAC as well as mixed-ADC system was also considered in \cite{zhang2019mixed} for relaying channels.
Bussgang decomposition was adopted in \cite{jacobsson2017quantized,li2017downlink, xu2019secure} to linearize the low-resolution DAC system to develop precoder and analyze system performance. 
Interestingly, it was shown in \cite{xu2019secure} that employing low-resolution DACs can offer more reliable secure communication depending on system configuration.
The AQNM was also used in \cite{ribeiro2018energy,dai2019achievable}.
In \cite{ribeiro2018energy}, numerical comparison among digital BF and hybrid analog and digital BF with fully-connected and partially-connected phase shifter networks  was provided.
In \cite{dai2019achievable}, using low-resolution ADCs and DACs  provided benefits in reducing power consumption while maintaining achievable rate in full duplex systems. 

The prior work on low-resolution ADCs and DACs discloses that using low-resolution quantizers can significantly reduce the power consumption at the BSs while maintaining desirable spectral efficiency.
Given the benefit of using low-resolution ADCs and DACs in the SE-EE trade-off, it is indispensable to consider coarse quantization systems for CoMP BF with massive antenna arrays.
However, the non-negligible quantization error that is a function of channels, beamformers and transmit power makes the CoMP problem more challenging to solve. 
Due to the quantization error, it is unknown whether previous findings in the prior work can still be valid in the low-resolution ADC and DAC systems. 
When it comes to the OFDM system, the quantization involves the OFDM modulations as well as BF and channels, which leads to highly complicated problems.
Therefore, comprehensive study on CoMP for massive MIMO systems with low-resolution ADCs and DACs is desirable.

\subsection{Contributions}

In this paper, we investigate joint BF and PC problems in coordinated multicell networks with BSs equipped with a large number of antenna arrays. 
We focus on coarse quantization systems where the BSs are equipped with low-resolution ADCs and DACs to achieve energy-efficient communications. 
Accordingly, the non-negligible quantization error which involves various system functions needs to be properly manipulated.
For tractability, we adopt the AQNM for modeling the quantization system.
The contributions are summarized as follows:
\begin{itemize}[leftmargin=*]
	\item We first formulate the minimum total transmit power problem  subject to individual SINR constraints for both DL and UL. 
	Then we prove the duality between the DL and UL problems under the coarse quantization systems by showing that the Lagrangian dual problem of the DL problem is equivalent to the UL problem with MMSE combiners.
	We further demonstrate that there is no duality gap, i.e., strong duality holds, by casting the DL problem into a standard second order conic problem and by showing strict feasibility with respect to the beamformer.
	\item Leveraging the strong duality, we propose a fixed point iterative algorithm to jointly solve the DL and UL problems.
    Using the properties of a standard function, we show that the algorithm converges to a unique optimal set of transmit powers for the UL problem.
    We further show that the optimal DL beamformers can be obtained by scaling the UL MMSE combiner that is design based on the optimal transmit powers.
    We also remark that the proposed algorithm can be implemented in a distributed manner with in-cell channel knowledge and without requiring explicit estimation of inter-cell channels. 
	\item Assuming homogeneous transmit powers and SINR constraints per cell, a deterministic algorithm is developed to provide a closed-form solution for the UL BF and PC problem. 
	To this end, we consider an MMSE equalizer and derive a lower bound of the minimum SINR for each cell.
	Then the solution is derived as a linear function of the SINR constraints and maximum eigenvalues of matrices that are composed of channels.
	\item We extend the CoMP BF and PC problem to a wideband OFDM system.
	We first derive  DL and UL system models by incorporating the coarse quantization effect into the OFDM modulation.
	Then we formulate the minimum total transmit power problems for UL and DL to find the optimal beamformer and transmit power for each user and subcarrier subject to the SINR constraints for each user and subcarrier.
	Manipulating the quantization error that is intertwined with not only the channels, beamformers, and transmit power but also the OFDM modulation, we show that the strong duality also holds for the wideband OFDM systems and the similar results as the narrowband system can be applied.
	\item Simulation results validate the derived theoretical results and demonstrate that the proposed iterative CoMP algorithm achieves the target SINR.
	The algorithm also outperforms a conventional per-cell based method in terms of accuracy and minimizing total transmit power. 
    In addition, the deterministic approach whose total transmit power lies between these two methods in the medium-to-high target SINR range show a reasonable trade-off between transmit power and achieved SINR.
\end{itemize}

{\it Notation}: $\bf{A}$ is a matrix and $\bf{a}$ is a column vector. 
$\mathbf{A}^{H}$ and $\mathbf{A}^T$  denote conjugate transpose and transpose. 
$[{\bf A}]_{i,:}$ and $ \mathbf{a}_i$ indicate the $i$th row and column vectors of $\bf A$. 
We denote $a_{i,j}$ as the $\{i,j\}$th element of $\bf A$ and $a_{i}$ as the $i$th element of $\bf a$. 
$\mathcal{CN}(\mu, \sigma^2)$ is a complex Gaussian distribution with mean $\mu$ and variance $\sigma^2$. 
The diagonal matrix $\rm diag(\bf A)$ has $\{a_{i,i}\}$ at its diagonal entries, and $\rm diag (\bf a)$ or $\rm diag({\bf a}^T)$ has $\{a_i\}$ at its diagonal entries. 
A block diagonal matrix is presented as ${\rm blkdiag}({\bf A}_1, \dots,{\bf A}_{N})$. 
A block circulant matrix is denoted as ${\rm blkcirc}(\bA_1,\dots,\bA_N)$ with a first block-row of $\bA_1,\dots,\bA_N$.
${\rm eig}_{\rm M}(\bA)$ and ${\rm eig}_{\rm m}(\bA)$ denote the maximum and minimum eigenvalues of $\bA$, respectively.
We use $\rm vec({\bf A})$ to represent the vectorization operator.
${\bf I}_N$ is a $N\times N$ identity matrix and ${\bf 0}_N$ is a $N \times 1$ zero vector.
$\|\bf A\|$ represents L2 norm. 

\section{System Model}
\label{sec:sys_model}

\begin{figure}[!t]\centering
\includegraphics[scale = 0.45]{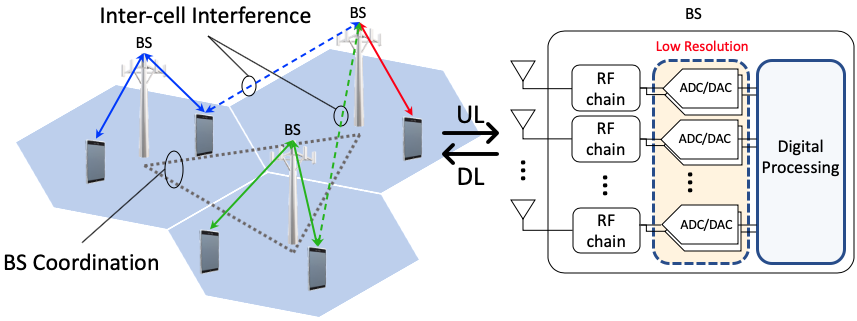}
\vspace{-1em}
\caption{Multicell multiuser-MIMO network which is incorporated with low-resolution ADCs and DACs at the BS.} 
\label{fig:system}
\end{figure}

We consider a multicell multiuser-MIMO network with $N_c$ cells, $N_{u}$ single-antenna users per cell.
Users are served by an associated BS with $N_b$ antennas ($N_b \gg N_u$), i.e., users in cell $i$ are served by a BS in cell $i$.
We assume that the BSs for all $N_c$ cells are equipped with low-resolution ADCs and DACs with equal bits, i.e., $b$-bit ADCs and DACs for all BSs, and they cooperate as shown in Fig.~\ref{fig:system}.
Time division multiplexing (TDD) is considered in the system.

\subsection{Uplink Narrowband System}

Each user $u$ in cell $i$ transmits signal $x^{\rm ul}_{i,u} = \sqrt{\lambda_{i,u}}s^{\rm ul}_{i,u}$ over a narrowband channel, where $\lambda_{i,u}$ and $s^{\rm ul}_{i,u}$ are transmit power and a symbol, respectively. 
The narrowband channel vector between user $u$ in cell $j$ and the BS in cell $i$ (BS$_i$) is represented as ${\bf h}_{i,j,u} \in \bbC^{N_b}$.
Then, the received baseband analog signal at BS$_i$ is expressed as
\begin{align}
    \label{eq:r_ul}
    {\bf r}^{\rm ul}_i 
    ={\bf H}_{i,i}{\bf x}^{\rm ul}_i + \sum_{j\neq i}^{N_c}{\bf H}_{i,j}{\bf x}^{\rm ul}_j + {\bf n}^{\rm ul}_i
    ={\bf H}_{i,i}{\pmb \Lambda}^{1/2}_i{\bf s}^{\rm ul}_i + \sum_{j\neq i}^{N_c}{\bf H}_{i,j}{\pmb \Lambda}^{1/2}_j{\bf s}^{\rm ul}_j + {\bf n}^{\rm ul}_i
\end{align} 
where ${\bf H}_{i,j} \in \bbC^{N_b\times N_u}$ is the channel matrix between BS$_i$ and users in  cell $j$ whose $u$th column is $\bh_{i,j,u}$, ${\bf x}^{\rm ul}_i \in \bbC^{N_u}$ and ${\bf s}^{\rm ul}_i \in \bbC^{N_u}$ are the transmit signal and symbol vectors of the $N_u$ users in cell $i$, whose $u$th entries are $x^{\rm ul}_{i,u}$ and $s^{\rm ul}_{i,u}$, respectively, ${\pmb \Lambda}_i = {\rm diag}(\lambda_{i,1},\dots,\lambda_{i,N_u})$ is the the transmit power matrix of the users in cell $i$, and ${\bf n}^{\rm ul}_i \in \mathbb{C}^{N_b}$ is the additive white Gaussian noise (AWGN) vector at BS$_i$.
Throughout this paper, we consider a normalized variance for AWGN without loss of generality, i.e, $\bn^{\rm ul}_i \sim \mathcal{CN}({\bf 0},{\bf I}_{N_b})$.
We further consider that  ${\bf s}^{\rm ul}_i$ has a zero mean and unit variance.
We can rewrite the analog received signal \eqref{eq:r_ul} in a more compact form as
\begin{align}
    \nonumber
    {\bf r}_i = {\bf H}_i{\pmb \Lambda}^{1/2}{\bf s}^{\rm ul} + {\bf n}_i^{\rm ul}
\end{align} 
where ${\bf H}_i = [{\bf H}_{i,1},\dots,{\bf H}_{i,N_c}] \in \bbC^{N_b\times N_cN_u}$, ${\pmb \Lambda} = {\rm blkdiag}({\pmb \Lambda}_1,\dots, {\pmb \Lambda}_{N_c})\in \bbC^{N_cN_u\times N_cN_u}$, and ${\bf s}^{\rm ul} = [({\bf s}_1^{\rm ul})^{T},\dots,({\bf s}_{N_c}^{\rm ul})^{T}]^T\in \bbC^{N_cN_u}$.

We consider  that each ADC has $b$ quantization bits.
We adopt the AQNM~\cite{fletcher2007robust, orhan2015low} to obtain a linearized approximation of the quantization process derived from assuming a scalar MMSE quantizer.
Under the AQNM, the quantized signal vector can be given as~\cite{fletcher2007robust}
\begin{align} 
    \label{eq:rq_ul}
     \cQ(\br_i) \approx {\bf r}_{{\rm q},i} 
    = \alpha {\bf H}_{i,i}{\pmb \Lambda}^{1/2}_i{\bf s}_i^{\rm ul} +  \alpha \sum_{j\neq i}^{N_c}{\bf H}_{i,j}{\pmb \Lambda}^{1/2}_j{\bf s}_j^{\rm ul} +  \alpha {\bf n}_i^{\rm ul} + \bq_i^{\rm ul}
\end{align} 
where $\mathcal{Q}(\cdot)$ is an element-wise quantizer function applied to the real and imaginary parts.
The quantization gain $\alpha$ is a function of the number of ADC bits and defined as $\alpha = 1- \beta$, where $\beta= \frac{\mathbb{E}[|{r} - {r}_{{\rm q}}|^2]}{\mathbb{E}[|{r}|^2]}$ \cite{fletcher2007robust,fan2015uplink}.
Assuming that $\bs_i^{\rm ul}$ is Gaussian distributed, i.e.,  $\bs^{\rm ul}_i\sim\mathcal{CN}({\bf 0},{\bf I}_{N_u}), \forall i$, the values of $\beta$ are listed in Table 1 in \cite{fan2015uplink} for $b \leq 5$, and $\beta$ is approximated as $\beta \approx \frac{\pi\sqrt{3}}{2} 2^{-2b}$ for $b > 5$ \cite{gersho2012vector}. 
The quantization noise $\bq_i^{\rm ul}$ is uncorrelated with ${\bf r}_i$ \cite{fletcher2007robust} and considered to follow the complex Gaussian distribution with a zero mean and covariance of \cite{fletcher2007robust,orhan2015low}
\begin{align}
   \label{eq:cov_q_ul}
    \mathbf{C}_{\bq_i^{\rm ul}\bq_i^{\rm ul}} = \alpha (1-\alpha) \,{\rm diag}\big({\bH_i}{\pmb \Lambda}{\bf H}_{i}^H + \bI_{N_b}\big).
\end{align}


Once the received signals are quantized, they are combined with ${\bf F}_i$ at BS$_i$.
Then we have 
\begin{align}
    \nonumber
    {\bf y}^{\rm ul}_i  = {\bf F}_i^H {\bf r}_{{\rm q},i}
    =  \alpha {\bf F}_i^H  {\bf H}_{i,i}{\pmb \Lambda}^{1/2}_i{\bf s}^{\rm ul}_i +  \alpha \sum_{j\neq i}^{N_c}  {\bf F}_i^H  {\bf H}_{i,j}{\pmb \Lambda}^{1/2}_j{\bf s}^{\rm ul}_j +  \alpha {\bf F}_i^H  {\bf n}^{\rm ul}_i + {\bf F}_i^H \bq_i^{\rm ul}.
\end{align}
Accordingly, the quantized and combined received signal for user $u$ in cell $i$ is given as
\begin{align}
    \nonumber
    y^{\rm ul}_{i,u} &= \alpha \sqrt{\lambda_{i,u}} {\bf f}_{i,u}^H {\bf h}_{i,i,u} s^{\rm ul}_{i,u}\! +\! \alpha \sum_{v\neq u}^{N_u}\sqrt{\lambda_{i,v}}\bff_{i,u}^H\bh_{i,i,v}s^{\rm ul}_{i,v}\!+\! \alpha \sum_{\substack{ j \neq i\\ v}}^{N_c,N_u} \sqrt{\lambda_{j,v}} {\bf f}_{i,u}^H {\bf h}_{i,j,v} s^{\rm ul}_{j,v} \!+\! \alpha  {\bf f}_{i,u}^H {\bf n}^{\rm ul}_{i} \!+\!  {\bf f}_{i,u}^H\bq_i^{\rm ul}\\
    \nonumber
    &= \alpha \sqrt{\lambda_{i,u}} {\bf f}_{i,u}^H {\bf h}_{i,i,u} s^{\rm ul}_{i,u} + \alpha \sum_{(j,v) \neq (i, u)}^{(N_c, N_u)} \sqrt{\lambda_{j,v}} {\bf f}_{i,u}^H {\bf h}_{i,j,v} s^{\rm ul}_{j,v} + \alpha  {\bf f}_{i,u}^H {\bf n}^{\rm ul}_{i} +  {\bf f}_{i,u}^H\bq_i^{\rm ul}
\end{align}
where $\bff_{i,u}$ is the $u$th column of $\bF_i$.

\subsection{Downlink Narrowband System}

Similarly to the UL quantized signals, the transmit signal vector quantized at low-resolution DACs of  BS$_i$ with a precoder ${\bf W}_i \in \bbC^{N_b\times N_u}$ is expressed as ${\bf x}^{\rm dl}_i = \alpha {\bf W}_i{\bf s}^{\rm dl}_i + \bq^{\rm dl}_i \in \bbC^{N_b}$ \cite{ribeiro2018energy,zhang2018mixed,dai2019achievable}, where ${\bf s}^{\rm dl}_i \sim\mathcal{CN}({\bf 0},{\bf I}_{N_u})$ denotes the transmit symbol vector for the $N_u$ users in cell $i$, and $\bq^{\rm dl}_i \in \bbC^{N_b}$ is a quantization noise vector with a covariance \cite{dai2019achievable}
\begin{align}
    \label{eq:Cqq_dl}
    {\bf C}_{\bq^{\rm dl}_i\bq^{\rm dl}_i} = \alpha (1-\alpha) {\rm diag}({\bf W}_i{\bf W}_i^H).
\end{align}
The same assumptions are made for the quantization as the UL system and $\alpha$ is also identical to the one in the UL system with the equal quantization resolution as the ADCs.
Under TDD, the channel vector between BS$_j$ and user $u$ in cell $i$ is $\bh^H_{j,i,u}$. 
The received signal at user $u$ is
\begin{align}
    \nonumber
    y^{\rm dl}_{i,u} = \alpha {\bf h}_{i,i,u}^H {\bf w}_{i,u} {s}_{i,u}^{\rm dl} + \alpha \sum_{(j,v)\neq (i,u)}^{(N_c,N_u)} {\bf h}_{j,i,v}^H  {\bf w}_{j,v} {s}^{\rm dl}_{j,v} + \sum_{j=1}^{N_c} {\bf h}_{j,i,u}^H \bq_j^{\rm dl} +  n^{\rm dl}_{i,u}
\end{align}
where $\bw_{i,u}$ is the $u$th column of $\bW_i$ and $n^{\rm dl}_{i,u}$ is the AWGN distributed as $n^{\rm dl}_{i,u}\sim\mathcal{CN}(0, 1)$.

\section{Uplink and Downlink Joint Beamforming and Power Control}
In this section, we formulate transmit power minimization problems for the UL and DL systems subject to given SINR constraints and propose algorithms that solve the problems.
In this paper, we assume that the problems are feasible.
First, the UL problem is formulated to minimize the transmit power of the users in $N_c$ cells with an individual user SINR constraint as
\begin{align}
    \label{eq:problem_ul}
    \cP1:\qquad \min_{{\bf f}_{i,u}, \lambda_{i,u}, \forall i,u} \sum_{i,u}^{N_c,N_u} \lambda_{i,u} \qquad \text{s.t. }\  \max_{{\bf f}_{i,u}} \Gamma^{\rm ul}_{i,u} \geq \gamma_{i,u},\ \forall \, i,u
\end{align}
where $\Gamma^{\rm ul}_{i,u}$ is the UL SINR of user $u$ in cell $i$, which is computed as
\begin{align}
    \label{eq:sinr_ul}
    \Gamma^{\rm ul}_{i,u} & = \frac{\alpha^2 \lambda_{i,u} |{\bff}_{i,u}^H {\bf h}_{i,i,u}|^2}{\alpha^2 \sum_{(j,v) \neq (i,u) }^{(N_c,N_u)} \lambda_{j,v} |{\bff}_{i,u}^H {\bf h}_{i,j,v}|^2  +  \alpha^2 \|{\bff}_{i,u}\|^2 + {\bff}_{i,u}^H {\bf C}_{\bq_i\bq_i}{\bff}_{i,u}}.
\end{align}
Unlike the perfect quantization system (no quantization error), $\Gamma^{\rm ul}_{i,u}$ has the additional term associated with quantization error, ${\bff}_{i,u}^H {\bf C}_{\bq_i\bq_i}{\bff}_{i,u}$, which is a function of the channel and the transmit power $\lambda_{i,u}$.
In addition, it is also involved with the combiner $\bff_{i,u}$. 
Accordingly, the effect of coarse quantization needs to be incorporated when solving  $\cP1$.

Now the DL problem is formulated to minimize the transmit power of the BSs in $N_c$ cells with an individual user SINR constraint as
 \begin{align}
    \label{eq:problem_dl}
    \cP2: \qquad \min_{{\bf w}_{i,u}, \forall i,u}\   \alpha\sum_{i,u}^{N_c,N_u} {\bf w}_{i,u}^H{\bf w}_{i,u}\qquad\text{subject to }\  \Gamma^{\rm dl}_{i,u} \geq \gamma_{i,u}, \quad \forall \, i,u
\end{align}
where
\begin{align}
    \label{eq:sinr_dl}
    \Gamma^{\rm dl}_{i,u} = \frac{\alpha^2 |{\bf w}_{i,u}^H {\bf h}_{i,i,u}|^2} { \alpha^2 \sum_{(j,v) \neq (i,u)}^{(N_c,N_u)}|{\bf w}_{j,v}^H {\bf h}_{j,i,u}|^2  + \sum_{j=1}^{N_c}{\bf h}_{j,i,u}^H {\bf C}_{\bq^{\rm dl}_j\bq^{\rm dl}_j} {\bf h}_{j,i,u} + 1}.
\end{align}
Note that $\alpha$ in the objective function is a fixed scalar which does not change the solution of $\cP2$.
The solution of $\cP2$ also needs to incorporate the effect of the coarse quantization, i.e., quantization noise covariance $\bC_{\bq^{\rm dl}_j\bq^{\rm dl}_j}$ as it is a function of $\bW_j$ and involved with channels ${\bf h}_{j,i,u}$.

\subsection{Uplink and Downlink Duality}

In this subsection, we extend the duality of the UL and DL power minimization problems for infinite-resolution quantizer systems \cite{dahrouj2010coordinated} to low-resolution quantizer systems by incorporating the quantization error terms. 
Exploiting the duality, we propose an iterative algorithm based on the fixed-point iteration \cite{yates1995framework} to solve both the UL and DL problems and further prove optimality and convergence of the algorithm.

\begin{theorem}[Duality]
    \label{thm:duality}
    The uplink transmit power minimization problem $\cP1$ in \eqref{eq:problem_ul} is equivalent to a Lagrangian dual problem of the downlink transmit minimization problem $\cP2$ in \eqref{eq:problem_dl}.
\begin{proof}
    The SINR constraints of $\cP1$ can be simplified by applying MMSE equalizers ${\bF}_{i}$ that maximize the SINR.
    Let ${\bf z}_{i,u}$ be the interference-plus-noise term of the UL quantized signal in \eqref{eq:rq_ul} whose covariance matrix is expressed as
    \begin{align}
        \nonumber
        {\bf C}_{{\bf z}_{i,u}{\bf z}_{i,u}} &= \alpha^2 \sum_{(j,v)\neq (i,u)} \lambda_{j,v} {\bf h}_{i,j,v}{\bf h}_{i,j,v}^H + \alpha^2 {\bf I}_{N_b} + \alpha(1-\alpha){\rm diag}({\bf H}_i {\pmb \Lambda} {\bf H}_i^H + {\bf I}_{N_b})\\
        \nonumber
        & = \alpha^2 \sum_{(j,v)\neq (i,u)} \lambda_{j,v} {\bf h}_{i,j,v}{\bf h}_{i,j,v}^H + \alpha {\bf I}_{N_b} + \alpha(1-\alpha){\rm diag}({\bf H}_i {\pmb \Lambda} {\bf H}_i^H).
    \end{align}
    Then, the MMSE equalizer ${\bf f}_{i,u}$ can be given as \cite{tse2005fundamentals}
    \begin{align}
        \label{eq:MMSE}
        {\bf f}_{i,u} = {\bf C}_{{\bf z}_{i,u}{\bf z}_{i,u}}^{-1} {\bf h}_{i,i,u}.
    \end{align}
    Applying \eqref{eq:MMSE} to the UL SINR in \eqref{eq:sinr_ul}, the constraints in $\cP1$ become  $\alpha^2 \lambda_{i,u} {\bf h}_{i,i,u}^H {\bf C}_{{\bf z}_{i,u}}^{-1} {\bf h}_{i,i,u} \geq \gamma_{i,u}$.
    Then we multiply both sides with $\bh_{i,i,u}^H\bh_{i,i,u}$ as
    \begin{align}
        \nonumber
        \alpha^2 \lambda_{i,u} {\bf h}_{i,i,u}^H{\bf h}_{i,i,u} {\bf h}_{i,i,u}^H {\bf C}_{{\bf z}_{i,u}{\bf z}_{i,u}}^{-1} {\bf h}_{i,i,u} &\geq \gamma_{i,u}{\bf h}_{i,i,u}^H{\bf h}_{i,i,u}\\
        \label{eq:duality_pf}
        {\bf h}_{i,i,u}^H(\alpha^2 \lambda_{i,u} {\bf h}_{i,i,u} {\bf h}_{i,i,u}^H {\bf C}_{{\bf z}_{i,u}{\bf z}_{i,u}}^{-1}-\gamma_{i,u} {\bf I}_{N_b}) {\bf h}_{i,i,u} &\geq 0
    \end{align}
    To satisfy \eqref{eq:duality_pf}, we need $\alpha^2 \lambda_{i,u} {\bf h}_{i,i,u} {\bf h}_{i,i,u}^H {\bf C}_{{\bf z}_{i,u}{\bf z}_{i,u}}^{-1}-\gamma_{i,u} {\bf I}_{N_b} \succeq 0$. 
    Rearranging this condition, we can rewrite $\cP1$ as 
    \begin{gather}
        \label{eq:problem_ul_simple}
        \min_{\lambda_{i,u}} \sum_{i,u} \lambda_{i,u}\quad \text{s.t. }\ {\bf K}_{i}(\pmb\Lambda) \preceq \alpha \bigg(1 + \frac{1}{\gamma_{i,u}}\bigg)\lambda_{i,u}  {\bf h}_{i,i,u}{\bf h}_{i,i,u}^H, \quad \forall i,u.
    \end{gather}
    where 
    \begin{align}
        \nonumber
        {\bf K}_{i}(\pmb \Lambda) = {\bf I}_{N_b}  + \alpha \sum_{j,v} \lambda_{j,v}{\bf h}_{i,j,v}{\bf h}_{i,j,v}^H + (1-\alpha) {\rm diag}\big({\bf H}_i {\pmb \Lambda} {\bf H}_i^H\big).
    \end{align}
    
    Now, we prove the duality between $\cP1$ and $\cP2$ by managing the quantization error term and by showing that the Lagrangian dual problem of $\cP2$ is equivalent to \eqref{eq:problem_ul_simple}.
    The Lagrangian of $\cP2$ is given as
    \begin{align}
        \nonumber
        \mathcal{L}({\bf w}_{i,u}, \mu_{i,u}) = & \sum_{i,u} \alpha {\bf w}_{i,u}^H{\bf w}_{i,u} - \sum_{i,u} \mu_{i,u} \Bigg( \alpha^2 \frac{|{\bf w}^H_{i,u} {\bf h}_{i,i,u}|^2}{\gamma_{i,u}} - \alpha^2 \sum_{v \neq u} {| {\bf w}_{i,v}^H {\bf h}_{i,i,u}|^2}   \\    \label{eq:lagrangian}
        &- \alpha^2 \sum_{\substack{j \neq i\\v}}|{\bf w}_{j,v}^H {\bf h}_{j,i,u}|^2  + \alpha (1-\alpha) \sum_{j}{\bf h}_{j,i,u}^H  {\rm diag}\big({\bf W}_j{\bf W}_j^H\big) {\bf h}_{j,i,u} + 1\Bigg) 
    \end{align}
    where $\mu_{i,u}$ is a Lagrangian multiplier.
    Rearranging and rewriting \eqref{eq:lagrangian}, the Lagrangian becomes
    \begin{align}
        \nonumber
        \mathcal{L}({\bf w}_{i,u}, &\mu_{i,u}) = \sum_{i,u}\mu_{i,u} + \alpha \sum_{i,u} {\bf w}_{i,u}^H \Bigg({\bf I}_{N_b}  - \alpha \bigg(1 + \frac{1}{\gamma_{i,u}}\bigg)\mu_{i,u}  {\bf h}_{i,i,u}{\bf h}_{i,i,u}^H \\ 
        \label{eq:lagrangian_pf1}
         & + \alpha \sum_{j,v} \mu_{j,v}{\bf h}_{i,j,v}{\bf h}_{i,j,v}^H \Bigg){\bf w}_{i,u} + \alpha (1-\alpha)  \sum_{i,u} \mu_{i,u}\sum_{j}{\bf h}_{j,i,u}^H  {\rm diag}({\bf W}_j{\bf W}_j^H) {\bf h}_{j,i,u}.
    \end{align}
    We need to rewrite the quantization error term in \eqref{eq:lagrangian_pf1} to manipulate $\bW_j$ in the diagonal matrix.
    Let $\bM_i = {\rm diag}(\mu_{i,1},\dots,\mu_{i,N_u})$ and ${\bM} = {\rm blkdiag}({\bM}_1,\dots, {\bM}_{N_c})\in \bbC^{N_cN_u\times N_cN_u}$.
    Changing the indices of $\sum_{i,u} \mu_{i,u}\sum_{j}{\bf h}_{j,i,u}^H  {\rm diag}({\bf W}_j{\bf W}_j^H) {\bf h}_{j,i,u}$ from $(i,u,j)$ to $(j,v,i)$, we have
    \begin{align}
        \nonumber
        \sum_{j,v}^{N_c,N_u} \!\!\mu_{j,v}\!\sum_{i}^{N_c}\!{\bf h}_{i,j,v}^H  {\rm diag}\big({\bf W}_i{\bf W}_i^H\big) {\bf h}_{i,j,v}\!&=\!\sum_{j,v}^{N_c,N_u} \mu_{j,v} \sum_{i,n}^{N_c,N_b} |h_{i,j,v,n}|^2 \sum_u^{N_u}|w_{i,u,n}|^2\\ 
        \nonumber
        &=\!\!\!\sum_{i,u}^{N_c,N_u}\!\!\bw^H_{i,u}{\rm diag}\Big(\!\sum_{j,v}^{N_c,N_u} \!\!\mu_{j,v}|h_{i,j,v,1}|^2,\dots,\!\sum_{j,v}^{N_c,N_u} \!\!\mu_{j,v}|h_{i,j,v,N_b}|^2\!\Big)\bw_{i,u} \\ 
        \label{eq:lagrangian_pf2} 
        &= \!\sum_{i,u}^{N_c,N_u} {\bf w}_{i,u}^H {\rm diag}({\bf H}_i {\pmb \bM} {\bf H}_i^H){\bf w}_{i,u},
    \end{align}
    where $h_{i,j,v,n}$ and $w_{i,u,n}$ are the $n$th entries of $\bh_{i,j,v}$ and  $\bw_{i,u}$, respectively, and ${\bf H}_i \!=\! [{\bf H}_{i,1},\dots,{\bf H}_{i,N_c}]$ as defined earlier.
    Applying \eqref{eq:lagrangian_pf2} to \eqref{eq:lagrangian_pf1}, the Lagrangian becomes
    \begin{align}
        \nonumber
        \mathcal{L}({\bf w}_{i,u}, \mu_{i,u}) =\, \sum_{i,u}\mu_{i,u} + \alpha& \sum_{i,u}  {\bf w}_{i,u}^H \Bigg({\bf I}_{N_b}  - \alpha \bigg(1 + \frac{1}{\gamma_{i,u}}\bigg)\mu_{i,u} {\bf h}_{i,i,u}{\bf h}_{i,i,u}^H \\ 
        \label{eq:lagrangian_pf3}
        & + \alpha \sum_{j,v} \mu_{j,v}{\bf h}_{i,j,v}{\bf h}_{i,j,v}^H + (1-\alpha) {\rm diag}\big({\bf H}_i {\bM} {\bf H}_i^H\big) \Bigg){\bf w}_{i,u}.
    \end{align}
    Let the dual objective function $g(\mu_{i,u}) = \min_{{\bf w}_{i,u}} \mathcal{L}({\bf w}_{i,u}, \mu_{i,u})$.  
    To prevent an unbounded solution, we need ${\bf I}_{N_b} - \alpha \Big(1 + \frac{1}{\gamma_{i,u}}\Big)\mu_{i,u}  {\bf h}_{i,i,u}{\bf h}_{i,i,u}^H + \alpha \sum_{j,v} \mu_{j,v}{\bf h}_{i,j,v}{\bf h}_{i,j,v}^H + (1-\alpha) {\rm diag}({\bf H}_i {\bM} {\bf H}_i^H) \succeq 0$.
    Accordingly, the Lagrangian dual problem of $\cP2$ in \eqref{eq:problem_dl} becomes equivalent to
    \begin{gather}
        \label{eq:lagrangian_pf3_2}
        \max_{\mu_{i,u}} \sum_{i,u}^{N_c,N_u} \mu_{i,u}\quad {\text{s.t. }}\ {\bf K}_{i}(\bM) \succeq \alpha \bigg(1 + \frac{1}{\gamma_{i,u}}\bigg)\mu_{i,u}  {\bf h}_{i,i,u}{\bf h}_{i,i,u}^H  , \quad \forall \, i,u
    \end{gather}
    where $ {\bf K}_{i}(\bM) = {\bf I}_{N_b}  + \alpha \sum_{j,v} \mu_{j,v}{\bf h}_{i,j,v}{\bf h}_{i,j,v}^H + (1-\alpha) {\rm diag}\big({\bf H}_i {\pmb \bM} {\bf H}_i^H\big)$.
    
    The dual problem in \eqref{eq:lagrangian_pf3_2} is equivalent to \eqref{eq:problem_ul_simple};
    although the Lagrangian dual problem of $\cP2$ in \eqref{eq:lagrangian_pf3_2} and the UL problem in \eqref{eq:problem_ul_simple} have the opposite objectives (max vs. min) with the reversed inequality in the constraints,  optimal solutions of $\cP1$ and the Lagrangian dual problem \eqref{eq:lagrangian_pf3_2} can be obtained with active constraints, and \eqref{eq:lagrangian_pf3_2} and \eqref{eq:problem_ul_simple} have the same optimal solutions with active constraints.
    Therefore, \eqref{eq:lagrangian_pf3_2} and \eqref{eq:problem_ul_simple} become equivalent  by replacing $\mu_{i,u}$ in \eqref{eq:lagrangian_pf3_2}  with $\lambda_{i,u}$, $\forall i,u$, i.e., the Lagrangian multiplier of $\cP2$, $\mu_{i,u}$, is indeed equivalent to the UL transmit power $\lambda_{i,u}$ in $\cP1$.
    This completes the proof for Theorem~\ref{thm:duality}.
\end{proof}
\end{theorem}
This result generalizes the UL-DL duality derived in \cite{dahrouj2010coordinated} to any quantization resolution since the $\cP1$ and $\cP2$ become equivalent to the UL and DL power minimization problem without quantization error, i.e., $b \to \infty \ (\text{equivalently, } \alpha \to 1)$.
To propose an algorithm which solves $\cP1$ and $\cP2$, and to prove its optimality, we first show strong duality between $\cP1$ and $\cP2$.


\begin{corollary}[Strong duality]
    \label{cor:strong_duality}
    Strong duality holds for $\cP2$ and its Lagrangian dual problem.
    \begin{proof}
    See Appendix~\ref{appx:strong_duality}.
    \end{proof}
\end{corollary}

\subsection{Distributed Iterative Algorithm}
\label{subsec:algorithm}

In this subsection, we characterize solutions by exploiting the strong duality between $\cP1$ and $\cP2$, and develop an iterative algorithm that finds the solutions for $\cP1$ and $\cP2$ simultaneously.
\begin{corollary}
    \label{cor:solution}
    The optimal transmit power for the uplink minimization problem \eqref{eq:problem_ul} is derived as
    \begin{align}
        \label{eq:solution}
        \lambda_{i,u} = \frac{1}{\alpha \Big(1+\frac{1}{\gamma_{i,u}}\Big){\bf h}_{i,i,u}^H {\bf K}_i^{-1}({\pmb \Lambda}) {\bf h}_{i,i,u}}
    \end{align}
    where $ {\bf K}_{i}({\pmb \Lambda}) = {\bf I}_{N_b}  + \alpha \sum_{j,v} \lambda_{j,v}{\bf h}_{i,j,v}{\bf h}_{i,j,v}^H + (1-\alpha) {\rm diag}({\bf H}_i {\pmb \Lambda} {\bf H}_i^H)$ 
    with the MMSE receiver given as
    \begin{align}
        \label{eq:MMSE2}
        {\bf f}_{i,u} =  \bigg( \alpha^2 \sum_{(j,v)\neq (i,u)} \lambda_{j,v} {\bf h}_{i,j,v}{\bf h}_{i,j,v}^H + \alpha {\bf I}_{N_b} + \alpha(1-\alpha){\rm diag}({\bf H}_i {\pmb \Lambda} {\bf H}_i^H)\bigg)^{-1}{\bf h}_{i,i,u}.
    \end{align}
\begin{proof}
    Here we use $\lambda_{i,u}$ instead of $\mu_{i,u}$ since we showed that they are equivalent.
    The derivative of the Lagrangian \eqref{eq:lagrangian_pf3} with respect to ${\bf w}_{i,u}$ is given as
    \begin{align}
        \nonumber
        \frac{\partial \mathcal{L}({\bf w}_{i,u}, \lambda_{i,u})}{\partial {\bf w}_{i,u}} = 2\alpha \Bigg({\bf I}_{N_b}  - \alpha \bigg(1 + \frac{1}{\gamma_{i,u}}\bigg)\lambda_{i,u}  {\bf h}_{i,i,u}{\bf h}_{i,i,u}^H &+ \alpha \sum_{j,v} \lambda_{j,v}{\bf h}_{i,j,v}{\bf h}_{i,j,v}^H \\
        \label{eq:zero_derivative}
         & + (1-\alpha) {\rm diag}({\bf H}_i {\pmb \Lambda} {\bf H}_i^H) \Bigg){\bf w}_{i,u}.
    \end{align}
    Setting \eqref{eq:zero_derivative} equal to zero, we derive \eqref{eq:solution}. 
    Accordingly, it is the Lagrangian multiplier that satisfies the stationarity condition. 
    In addition, at the optimal solution, all the constraints in $\cP2$ are active, which satisfies the complementary slackness condition.
    Therefore, \eqref{eq:solution} is the optimal Lagrangian multiplier, equivalently, optimal transmit power for $\cP1$.
\end{proof}
\end{corollary}

The solution in \eqref{eq:solution}, however, is a function of all transmit powers including itself.
Hence the solution does not fully solve the problem; we develop an algorithm to find an optimal set of transmit power by utilizing the solution.
Once we find the optimal transmit power, we can compute the MMSE combiner $\bF_i$ based on the transmit power. 
In addition, we show the linear relationship between the optimal UL MMSE combiner and the optimal DL precoder; the optimal DL precoder is a scaled version of the UL MMSE combiner.

\begin{corollary}[DL precoder]
    \label{cor:DL_precoder}
    With the carefully designed scaling factor, an optimal downlink precoder can be linearly proportional to the uplink MMSE receiver, i.e.,  $\bw_{i,u} = \sqrt{\tau_{i,u}}\bff_{i,u} \;\forall i, u$, and $\tau_{i,u}$ is derived as $\btau = \bSigma^{-1}{\bf 1}$, where ${\bf 1}$ is a $N_uN_c \times 1$ column vector with entries of all ones, $\btau = [\btau_{1}^T, \btau_{2}^T, \cdots, \btau_{N_c}^T ]^T$ with  $\btau_{i}^T = [\tau_{i,1}, \tau_{i,2}, \cdots, \tau_{i,N_u}]^T$, and  
    \begin{equation}
    \bSigma = 
    \begin{pmatrix}
    \bSigma_{1,1} & \bSigma_{1,2} & \cdots & \bSigma_{1,N_c} \\
    \bSigma_{2,1} & \bSigma_{2,2} & \cdots & \bSigma_{2,N_c} \\
    \vdots  & \vdots  & \ddots & \vdots  \\
    \bSigma_{N_c,1} & \bSigma_{N_c,2} & \cdots & \bSigma_{N_c,N_c}
    \end{pmatrix}.
    \end{equation}
    Each element of $\bSigma_{i,j}\in\bbR^{N_u \times N_u}$ is defined as \eqref{eq:DLprecoder}
    \begin{equation} 
        \label{eq:DLprecoder}
        [\bSigma_{i,j}]_{u,v} = 
        \begin{cases}
        \frac{\alpha^2}{\gamma_{i,u}} |\bff_{i,u}^H {\bf h}_{i,i,u}|^2 - \alpha(1-\alpha)\bff_{i,u}^H{\rm diag}(\bh_{i,i,u}\bh_{i,i,u}^H)\bff_{i,u} & \text{if } i=j \text{ and } u=v, \\
        -  \alpha^2 | \bff_{j,v}^H {\bf h}_{j,i,u}|^2 - \alpha(1-\alpha)\bff_{j,v}^H{\rm diag}(\bh_{j,i,u}\bh_{j,i,u}^H)\bff_{j,v} & \text{otherwise.}
        \end{cases}
    \end{equation}
    \begin{proof}
        See Appendix~\ref{appx:DL_precoder}.
    \end{proof}
\end{corollary}

Now, we use an iterative standard PC algorithm \cite{yates1995framework,wiesel2005linear,dahrouj2010coordinated} to find the optimal UL transmit power by exploiting \eqref{eq:solution}, which allows us to compute the optimal UL MMSE combiner and DL precoder;
let $\lambda_{i,u}^{(n)}$ be the UL transmit power at $n$th iteration.
The algorithm is  as follows:
\begin{description}
    \item[Step 1.] Initialize $\lambda_{i,u}^{(0)}$, $\forall i,u$.
    \item[Step 2.] Iteratively update the transmit power $\lambda_{i,u}^{(n+1)}$ until converges, using \eqref{eq:solution} as
        \begin{align}
            \label{eq:solution2}
            \lambda_{i,u}^{(n+1)} = \frac{1}{\alpha \Big(1+\frac{1}{\gamma_{i,u}}\Big){\bf h}_{i,i,u}^H {\bf K}_i^{-1}(\pmb \Lambda^{(n)}) {\bf h}_{i,i,u}}, \quad \forall i,u.
        \end{align}
    \item[Step 3.] Find the UL MMSE combiner ${\bf f}_{i,u}$ in \eqref{eq:MMSE2} with $\lambda_{i,u}$ obtained from the Step 1 and 2.
    \item[Step 4.] Compute the DL precoder $\bw_{i,u}$ based on Corollary~\ref{cor:DL_precoder}.
\end{description}
As remarked in \cite{dahrouj2010coordinated}, ${\bf K}_{i}$ is a covariance matrix of received signals which may be estimated locally at each BS$_i$, the fixed-point iteration in Step 2 for the optimal UL transmit power only requires the user channel information in the associated cell at the BS without the need for the explicit out-of-cell channel knowledge. 
In addition, the scaling coefficient $\tau_{i,u}$ for each user can be considered as a DL transmit power on the effective channel that achieves the target SINR. 
According to \cite{foschini1993simple}, the transmit power (equivalently,  $\tau_{i,u}$) can be obtained using a per-user power update algorithm, whose convergence is guaranteed \cite{yates1995framework}; each step of the algorithm computes $\tau_{i,u}$ that satisfies its target SINR while assuming other $\tau_{i',u'}$'s are fixed. 
Therefore, the proposed algorithm can be implemented in a distributed manner.
\begin{corollary}[Convergence]
    \label{cor:convergence}
    For any initial points $\lambda_{i,u}^{(0)}$, $\forall i,u$, the proposed fixed-point iterative algorithm converges to an unique fixed point at which total transmit power is minimized.
    \begin{proof}
        The proof is based on the standard function approach \cite{yates1995framework}. 
    Let us rewrite \eqref{eq:solution2} as $\lambda_{i,u}^{(n+1)} = \mathcal{F}_{i,u}(\pmb \Lambda^{(n)})$. 
    We need to show that $\mathcal{F}_{i,u}(\pmb \lambda)$ is a standard function which satisfies the followings:
    \begin{itemize}
        \item (positivity) If $\lambda_{i,u} \geq 0$ $\forall i,u$, then $\mathcal{F}_{i,u}({\pmb \Lambda}) > 0$.
        \item (monotonicity) If $\lambda_{i,u} \geq \lambda_{i,u}' \forall i,u$, then $\mathcal{F}_{i,u}({\pmb \Lambda}) \geq \mathcal{F}_{i,u}({\pmb \Lambda}')$.  
        \item (scalability) For $\rho > 1$, $\rho \mathcal{F}_{i,u}({\pmb \Lambda}) > \mathcal{F}_{i,u}(\rho{\pmb \Lambda})$.
    \end{itemize}
    It can be shown that $\mathcal{F}_{i,u}(\pmb \Lambda^{(n)})$ satisfies the properties by carefully following the proof in Appendix II in \cite{wiesel2005linear}.
    \end{proof}
\end{corollary}
Therefore, the fixed-point iteration in Step 2 always converges to an unique fixed point that is the optimal transmit power, and the optimal solutions for $\cP1$ and $\cP2$ can be obtained.

\subsection{Deterministic Solution for Homogeneous Transmit Power and SINR Constraint per Cell}

In this subsection, we derive a deterministic transmit power solution for a special case in which transmit powers and SINR constraints are homogeneous within each cell for UL, i.e., $\lambda_{i,u} = \lambda_i$ and $\gamma_{i,u} =\gamma_{i}$, $\forall u$.
We solve this problem by forcing the minimum SINR to satisfy the SINR constraint; $\min_u \Gamma_{i,u} \geq \gamma_i$, $\forall i,u$, and by relaxing the problem with the lower bound of the minimum SINR. 
With the MMSE equalizer $\bF_i$, the matrix of MSE for  UL in cell $i$ becomes
\begin{align}
    \nonumber
    \bE_{i}^{\rm mmse} = \left(\alpha^2\lambda_i\bH_{i,i}^H\left(\alpha^2 \sum_{j\neq i}^{N_c}\lambda_j\bH_{i,j}\bH_{i,j}^H+\alpha^2 \bI_{N_b} + \bC_{\bq_i^{\rm ul}\bq_i^{\rm ul}}\right)^{-1}\bH_{i,i}+\bI_{N_u}\right)^{-1}.
\end{align}
Accordingly, the SINR of user $u$ in cell $i$ can be expressed as $\Gamma_{i,u} = 1/[\bE^{\rm mmse}_i]_{u,u} -1$.
As shown in \cite{chen2007uplink}, the minimum SINR in cell $i$ is given as
\begin{align}
    \nonumber
    \min_u\Gamma_{i,u} &= \frac{1}{\max_u [\bE_{i}^{\rm mmse}]_{u,u}}-1\\
    \nonumber
    &\geq \frac{1}{{\rm eig}_{\rm M}\left(\bE_{i}^{\rm mmse}\right)}-1\\
    \label{eq:homo_proof1}
    &=\! {\rm eig}_{\rm m}\!\left(\alpha\lambda_i\bH_{i,i}^H\!\left(\alpha \sum_{j\neq i}^{N_c}\lambda_j\bH_{i,j}\bH_{i,j}^H\!+ \!\bI_{N_b}\! +\! (1\!-\!\alpha){\rm diag}(\bH_i\pmb \Lambda\bH_i^H)\right)^{-1}\!\!\bH_{i,i}\right).
\end{align}
Let $\bH_{i,i}^\dagger = (\bH_{i,i}^H\bH_{i,i})^{-1}\bH_{i,i}^H$, $A_{i,j}= {\rm eig}_{\rm M}\!\left(\!\bH_{i,i}^\dagger\bH_{i,j}\bH_{i,j}^H \bH_{i,i}^{\dagger H}\!\right)$, $B_i ={\rm eig}_{\rm M}\!\left(\!\!\big(\bH_{i,i}^H\bH_{i,i}\big)^{-1}\!\right)$, and $C_{i,j} = {\rm eig}_{\rm M}\!\left(\bH_{i,i}^\dagger{\rm diag}(\bH_{i,j}\bH_{i,j}^H)\bH_{i,i}^{\dagger H}\right)$.
Then \eqref{eq:homo_proof1} further becomes
\begin{align}
    \nonumber
    &\frac{\alpha\lambda_i}{{\rm eig}_{\rm M}\left(\bH_{i,i}^\dagger\left(\alpha\sum_{j\neq i}\lambda_j\bH_{i,j}\bH_{i,j}^H + \bI_{N_b} + (1-\alpha){\rm diag}(\bH_i\pmb\Lambda\bH_i^H)\right)\bH_{i,i}^{\dagger H}\right)}\\
    \nonumber
    &\stackrel{(a)}\geq \frac{\alpha\lambda_i}{\alpha\sum_{j\neq i}\lambda_j{\rm eig}_{\rm M}\!\left(\!\bH_{i,i}^\dagger\bH_{i,j}\bH_{i,j}^H \bH_{i,i}^{\dagger H}\!\right)\!+ \!{\rm eig}_{\rm M}\!\left(\bH_{i,i}^\dagger\bH_{i,i}^{\dagger H}\!\right)\! +\! (1\!-\!\alpha){\rm eig}_{\rm M}\!\left(\bH_{i,i}^\dagger{\rm diag}(\bH_i\pmb\Lambda\bH_i^H)\bH_{i,i}^{\dagger H}\right)}\\
    \label{eq:homo_proof2}
    & \stackrel{(b)} \geq \frac{\alpha\lambda_i}{\alpha\sum_{j\neq i}\lambda_j A_{i,j}+ B_i + (1\!-\!\alpha)\sum_j\lambda_j C_{i,j}}
\end{align}
where $(a)$ comes from Corollary~1 in \cite{chen2007uplink}, and $(b)$ is from ${\rm diag}(\bH_i\pmb\Lambda \bH_i^H) = \sum_j \lambda_j{\rm diag}(\bH_{i,j}\bH_{i,j}^H)$ due to $\pmb \Lambda_{i} = \lambda_i\bI_{N_u}$, $\forall i$ and Corollary~1 in \cite{chen2007uplink}. 

Setting \eqref{eq:homo_proof2} equal to $\gamma_i$ $\forall i$, we have the following linear equation:
\begin{align}
    \nonumber
    \pmb \lambda = \frac{1}{\alpha}\pmb\Gamma(\pmb \Omega\pmb \lambda + \bb)
\end{align}
where $\pmb \lambda = [\lambda_1,\dots,\lambda_{N_c}]^T$, $\pmb\Gamma = {\rm diag}(\gamma_1,\dots,\gamma_{N_c})$, $\bb = [B_1, \dots, B_{N_c}]^T$, and the $(i,j)$th element of $\pmb \Omega$ is given as
\begin{equation} 
    \nonumber
    \omega_{i,j} = 
    \begin{cases}
        (1-\alpha)C_{i,i} & \text{if } i=j  \\
       \alpha A_{i,j} + (1-\alpha)C_{i,j} & \text{otherwise.}
    \end{cases}
\end{equation}
Finally, the deterministic UL transmit power can be derived as
\begin{align}
    \label{eq:solution_deterministic}
    \pmb \lambda = \frac{1}{\alpha}(\bI_{N_c} - \frac{1}{\alpha}\pmb \Gamma \pmb\Omega)^{-1}\pmb \Gamma\bb.
\end{align}

We note that the deterministic solution in \eqref{eq:solution_deterministic} may have negative $\lambda_i$ when the target SINRs become high, i.e., the problem may easily become infeasible since the deterministic approach has a reduced feasible set by assuming homogeneous transmit powers per cell and by solving the  problem for the SINR lower bound.
We briefly introduce a possible approach to manage this issue.
Since the communication often operates in the interference-limited regime in the multicell system, changing the signs of all transmit powers only causes a marginal change in the SINR according to \eqref{eq:sinr_ul}. 
In this regard, if $\lambda_i <0 $, $\forall i$, we simply take the absolute value of $\lambda_i$ as a solution. 
If there exists $\lambda_i <0 $ only for a subset of the cells, we can set the largest $\lambda_i$ to zero and re-compute \eqref{eq:solution_deterministic} until we have $\lambda_i \geq 0$, $\forall i$, because the cell with the large $\lambda_i$ can be considered to have weak channels.
As a result, some of the cells can be assigned with zero transmit power. 
Then those cells can be scheduled in different time or frequency resources.

\section{Extension to Wideband OFDM Systems}

In this section, we extend the transmit power minimization problem to wideband OFDM systems under coarse quantization at the BSs.
To this end, we first need to derive signal models for multicell OFDM systems by taking into account quantization error coupled with OFDM modulations across all BSs.

\subsection{Uplink OFDM System with Low-resolution ADCs}
\label{subsec:ul_odfm}

Let $\bs^{\rm ul}_i(k) \in \bbC^{N_u}$ denote the vector of symbols of $N_u$ users in cell $i$ at subcarrier $k$, and let 
\begin{align}
    \nonumber
    \bu^{\rm ul}_i(k) = {\pmb \Lambda}_i(k)^{1/2}\bs^{\rm ul}_i(k),
\end{align} 
where ${\pmb \Lambda}_i(k) = {\rm diag}\big(\lambda_{i,1}(k),\dots,\lambda_{i,N_u}(k)\big)$ is the diagonal matrix of transmit power.
Let $\bx_i(k)^{\rm ul} \in \bC^{N_u}$ be the vector of OFDM symbols of $N_u$ users in cell $i$ at time $k$.  
We stack the OFDM symbol vectors as $\underline{\bx}_i^{\rm ul} = [\bx^{\rm ul}_i(0)^T,\dots,\bx^{\rm ul}_i(K-1)^T]^T \in \bbC^{KN_u}$.
Then $\underline{\bx}_i^{\rm ul}$ can be represented as
\begin{align}
    \nonumber
    \underline{\bx}_i^{\rm ul} &= (\bW_{\rm DFT}^H \otimes \bI_{N_u})\underline{\bu}^{\rm ul}_i = {\pmb \Psi}_{N_u}^H\underline{\pmb \Lambda}^{1/2}_i\underline{\bs}^{\rm ul}_i
\end{align}
where ${\pmb \Psi}_{N_u} = (\bW_{\rm DFT} \otimes \bI_{N_u})$, $\underline{\bu}_i^{\rm ul} = [\bu^{\rm ul}_i(0)^T,\dots,\bu^{\rm ul}_i(K-1)^T]^T$, $\underline{\bs}_i^{\rm ul} = [\bs^{\rm ul}_i(0)^T,\dots,\bs^{\rm ul}_i(K-1)^T]^T$, and $\underline{\pmb \Lambda}_i = {\rm blkdiag}\big({\pmb \Lambda}_i(0),\dots,{\pmb \Lambda}_i(K-1)\big)$.

Let $\br^{\rm ul}_i(k) \in \bbC^{N_b}$ be the received baseband analog signal at time $k$ after cyclic prefix (CP) removal at BS$_i$.
Staking for $K$-symbol time as $\underline{\br}^{\rm ul}_i = [\br^{\rm ul}_i(0)^T,\dots,\br^{\rm ul}_i(K-1)^T]^T \in \bbC^{KN_b}$, the stacked received baseband analog signals at BS$_i$ is expressed as
\begin{align}
    \nonumber
    \underline{\br}^{\rm ul}_i = \underline{\bH}_{i,i}\underline{\bx}^{\rm ul}_i + \sum_{j\neq i}^{N_c}\underline{\bH}_{i,j}\underline{\bx}^{\rm ul}_j + \underline{\bn}^{\rm ul}_i
\end{align}
where $\underline{\bH}_{i,j} = {\rm blkcirc}\big(\bH_{i,j,0},{\bf 0},\dots,{\bf 0},\bH_{i,j,L-1},\dots,\bH_{i,j,1}\big) \in \bbC^{KN_b\times KN_u}$ represents the block circulant channel matrix, $\bH_{i,j,\ell}$ denotes the time domain channel matrix between BS$_i$ and users in cell $j$ for $\ell$th tap, $L$ is the channel delay spread, and $\underline{\bn}^{\rm ul}_i$ is the stacked AWGN vector $\ubn^{\rm ul}_i= [\bn^{\rm ul}_i(0)^T,\dots,\bn^{\rm ul}_i(K-1)^T]^T\sim \cC\cN({\bf 0}, \bI_{KN_b})$.

The received signals are quantized and expressed under the AQNM as 
\begin{align}
    \nonumber
    \cQ(\underline{\br}^{\rm ul}_i) \approx \underline{\br}^{\rm ul}_{{\rm q},i} =  \alpha\underline{\bH}_{i,i}\underline{\bx}^{\rm ul}_i + \alpha\sum_{j\neq i}^{N_c}\underline{\bH}_{i,j}\underline{\bx}^{\rm ul}_j + \alpha\underline{\bn}^{\rm ul}_i + \underline{\bq}^{\rm ul}_i
\end{align}
where $\underline{\bq}^{\rm ul}_i = [\bq^{\rm ul}_i(0)^T,\dots,\bq^{\rm ul}_i(K-1)^T]^T \in \bbC^{KN_b} \sim \cC\cN({\bf 0}, \bC_{\underline{\bf q}^{\rm ul}_i\underline{\bf q}^{\rm ul}_i})$ is the stacked quantization noise vector for the received signal at BS$_i$, whose covariance matrix is \cite{fletcher2007robust} 
\begin{align}
    \label{eq:Cqq_ul_ofdm}
    \bC_{\underline{\bq}^{\rm ul}_i\underline{\bq}^{\rm ul}_i} = \alpha(1-\alpha) {\rm diag}\Big(\sum_{j=1}^{N_c}\underline{\bH}_{i,j}{\pmb \Psi}_{N_u}^H\underline{\pmb \Lambda}_j{\pmb \Psi}_{N_u}\underline{\bH}_{i,j}+\bI_{KN_b}\Big).
\end{align}
Now the quantized signals go through DFT operation and become
\begin{align}
    \nonumber
    \underline{\by}^{\rm ul}_i &= (\bW_{\rm DFT}\otimes \bI_{N_b})\underline{\br}^{\rm ul}_i \\
    \nonumber
    & = \alpha {\pmb \Psi}_{N_b} \underline{\bH}_{i,i}{\pmb \Psi}_{N_u}^H\underline{\pmb \Lambda}^{1/2}_i\underline{\bs}^{\rm ul}_i +  \alpha \sum_{j\neq i}{\pmb \Psi}_{N_b} \underline{\bH}_{i,j}{\pmb \Psi}_{N_u}^H\underline{\pmb \Lambda}^{1/2}_j\underline{\bs}^{\rm ul}_j + {\pmb \Psi}_{N_b} {\ubn}^{\rm ul}_i + {\pmb \Psi}_{N_b}{\ubq}^{\rm ul}_i\\
    \nonumber
    & = \alpha {\ubG}_{i,i}\,\underline{\pmb \Lambda}^{1/2}_i{\ubs}^{\rm ul}_i +  \alpha \sum_{j\neq i}{\ubG}_{i,j}\underline{\pmb \Lambda}^{1/2}_j{\ubs}^{\rm ul}_j +  \tilde{\ubn}^{\rm ul}_i + \tilde{\ubq}^{\rm ul}_i
\end{align}
where ${\pmb \Psi}_{N_b} = \bW_{\rm DFT}\otimes \bI_{N_b}$, $\ubG_{i,j} ={\pmb \Psi}_{N_b}\,\ubH_{i,j}{\pmb \Psi}_{N_u}^H = {\rm blkdiag}\big(\bG_{i,j}(0),\cdots,\bG_{i,j}(K-1)\big) \in \bbC^{KN_b \times KN_u}$ where $\bG_{i,j}(k) = \sum_{\ell = 0}^{L-1}\bH_{i,j,\ell} \,e^{-\frac{j2\pi k\ell}{K}}$ is the frequency domain UL channel matrix for subcarrier $k$ between BS$_i$ and users in cell $j$, $\tilde{\ubn}^{\rm ul}_i = [\tilde{\bn}^{\rm ul}_i(0)^T,\dots, \tilde{\bn}^{\rm ul}_i(K-1)^T]^T = {\pmb \Psi}_{N_b} {\ubn}^{\rm ul}_i$, and  $\tilde{\ubq}^{\rm ul}_i  = [\tilde{\bq}^{\rm ul}_i(0)^T,\dots, \tilde{\bq}^{\rm ul}_i(K-1)^T]^T= {\pmb \Psi}_{N_b} {\ubq}^{\rm ul}_i$.

The received signal at subcarrier $k$ is then given as
\begin{align}
    \label{eq:yk_ul_ofdm}
    \by^{\rm ul}_i(k) = \alpha\bG_{i,i}(k){\pmb \Lambda}_i(k)\bs^{\rm ul}_i(k) + \alpha\sum_{j\neq i}^{N_c}\bG_{i,j}(k){\pmb \Lambda}_j(k)\bs^{\rm ul}_j(k) + \alpha \tilde{\bn}^{\rm ul}_i(k) +\tilde{\bq}^{\rm ul}_i(k) 
\end{align}
and $\by^{\rm ul}_i(k)$ is combined with an equalizer $\bF_{i}(k)$.
The combined signal for user $u$ at subcarrier $k$ is now given as
\begin{align}
    \nonumber
    \bff_{i,u}^H(k)\by^{\rm ul}_i(k) =&\  \alpha\lambda_{i,u}^{1/2}(k)\bff_{i,u}^H(k)\bg_{i,i,u}(k)s^{\rm ul}_{i,u}(k)\\
    \nonumber
    &+ \alpha \!\!\!\sum_{(j,v)\neq (i,u)}^{N_c,N_u}\!\!\lambda_{j,v}^{1/2}(k)\bff_{i,u}^H(k)\bg_{i,j,v}(k)s^{\rm ul}_{i,v}(k)  + \alpha\bff_{i,u}^H(k)\tilde{\bn}_i(k) + \bff_{i,u}^H(k)\tilde{\bq}^{\rm ul}_i(k),
\end{align}
where $\bff_{i,u}^H(k)$ is the $u$th column of $\bF_i(k)$ and $\bg_{i,j,v}(k)$ is the $v$th column of $\bG_{i,j}(k)$.
We note that $\tilde{\ubn}^{\rm ul}_i \sim \cC\cN({\bf 0},\bI_{KN_b})$. 
The SINR for user $u$ in cell $i$ at subcarrier $k$ is computed accordingly as
\begin{align}
    \label{eq:sinr_ul_ofdm}
    {\Gamma}^{\rm ul}_{i,u}(k) = \frac{ \alpha^2\lambda_{i,u}(k)|\bff_{i,u}^H(k)\bg_{i,i,u}(k)|^2 }{ \alpha^2 \!\sum_{(j,v)\neq (i,u)}^{N_c,N_u}\!\lambda_{j,v}(k)|\bff_{i,u}^H(k)\bg_{i,j,v}(k)|^2  + \alpha^2 |\bff_{i,u}(k)|^2 + \bff_{i,u}^H(k)\bC_{\tilde{\bq}^{\rm ul}_i(k)\tilde{\bq}^{\rm ul}_i(k)}\bff_{i,u}(k) }.
\end{align}
Based on \eqref{eq:Cqq_ul_ofdm}, $\bC_{\tilde{\bq}_i(k)\tilde{\bq}_i(k)}$ is expressed as
\begin{align}
    \nonumber
    \bC_{\tilde{\bq}^{\rm ul}_i(k)\tilde{\bq}^{\rm ul}_i(k)} = \alpha(1-\alpha) {\pmb \Psi}_{N_b}(k){\rm diag}\Big(\sum_{j=1}^{N_c}\underline{\bH}_{i,j}{\pmb \Psi}_{N_u}^H\underline{\pmb \Lambda}_j{\pmb \Psi}_{N_u}\underline{\bH}_{i,j}+\bI_{KN_b}\Big){\pmb \Psi}_{N_b}^H(k)
\end{align}
where ${\pmb \Psi}_{N_b}(k) = \big([\bW_{{\rm DFT}}]_{k+1,:}\otimes \bI_{N_b}\big)$.
Finally, using \eqref{eq:sinr_ul_ofdm}, the UL OFDM transmit power minimization problem is formulated as 
\begin{align}
    \label{eq:problem_ul_ofdm}
   \cP3:\qquad \max_{{\bf f}_{i,u}(k), \lambda_{i,u}(k)} \sum_{i,u,k}\!\! \lambda_{i,u}(k) \qquad \text{s.t. }\  \max_{{\bf f}_{i,u}(k)} \Gamma^{\rm ul}_{i,u}(k) \geq \gamma_{i,u,k},\ \forall \, i,u,k.
\end{align}
In addition to all users in all cells, the maximization needs to be performed for all subcarriers.

\subsection{Downlink OFDM System with Low-resolution DACs}

The DL OFDM system with low-resolution DACs can be modeled by following similar steps as the UL OFDM system with low-resolution ADCs. 
Accordingly, we briefly explain the system model by pointing out the key differences such as precoding and DAC quantization, and definitions of symbols are the same as the ones used in Sec.~\ref{subsec:ul_odfm} unless mentioned otherwise.
Similarly to the UL OFDM system, the stacked OFDM symbol vector at BS$_i$ over $K$-symbol time, $ \ubx^{\rm dl}_i \in \bbC^{KN_b}$, is expressed as
\begin{align}
    \nonumber
    \ubx^{\rm dl}_i &= (\bW_{\rm DFT}^H\otimes \bI_{ N_b})\ubu^{\rm dl}_i = {\pmb \Psi}_{N_b}^H \ubW_i\ubs^{\rm dl}_i
\end{align}
where the block diagonal precoding matrix is $\ubW_i = {\rm blkdiag}\big(\bW_i(0),\dots,\bW_i(K-1)\big) \in \bbC^{KN_b\times KN_u}$.
Before being transmitted, $\ubx^{\rm dl}_i$ is quantized at the low-resolution DACs as \cite{fletcher2007robust, zhang2018mixed}
\begin{align}
    \nonumber
    \ubx^{\rm dl}_{{\rm q},i} = \alpha \ubx^{\rm dl}_i + \ubq^{\rm dl}_i
\end{align}
where $\ubq^{\rm dl}_i \sim \cC\cN({\bf 0}, \bC_{\underline{\bf q}^{\rm dl}_i\underline{\bf q}^{\rm dl}_i})$ is the stacked quantization noise vector at BS$_i$ and its covariance matrix is computed as \cite{fletcher2007robust}
\begin{align}
    \label{eq:Cqq_dl_ofdm}
    \bC_{\underline{\bf q}^{\rm dl}_i\underline{\bf q}^{\rm dl}_i} = \alpha(1-\alpha){\rm diag}\big({\pmb \Psi}_{N_b}^H\ubW_i\ubW_i^H{\pmb \Psi}_{N_b}\big).
\end{align}

After transmitting $\ubx^{\rm dl}_i $, $N_u$ users in cell $i$ receive signals from all BSs. 
Stacking over $K$ subcarriers after CP removal and DFT, the received signals at the users in cell $i$ becomes 
\begin{align}
    \nonumber
    \uby^{\rm dl}_i &= \alpha\ubG_{i,i}^H\ubW_i\ubs^{\rm dl}_i + \alpha\sum_{j\neq i}^{N_c}\ubG_{j,i}^H\ubW_j\ubs^{\rm dl}_j + \sum_{j=1}^{N_c}\ubG_{j,i}^H{\pmb \Psi}_{N_b}\ubq^{\rm dl}_j + {\pmb \Psi}_{N_u}\ubn_i^{\rm dl}\\
    \nonumber
    & =\alpha\ubG_{i,i}^H\ubW_i\ubs^{\rm dl}_i + \alpha\sum_{j\neq i}^{N_c}\ubG_{j,i}^H\ubW_j\ubs^{\rm dl}_j + \tilde{\ubq}^{\rm dl}_j + \tilde{\ubn}_i^{\rm dl} 
\end{align}
where $\tilde{\ubq}^{\rm dl}_j \!=\! \sum_{j=1}^{N_c}\ubG_{j,i}^H{\pmb \Psi}_{N_b}\ubq^{\rm dl}_j $ and $\tilde{\ubn}_i^{\rm dl} \! =\! {\pmb \Psi}_{N_u}\ubn_i^{\rm dl}$. 
Recall that $\ubG_{j,i}\!=\! {\rm blkdiag}(\bG_{j,i}(0),\!\cdots\!,\bG_{j,i}(K\!-\!1)) \in \bbC^{KN_b \times KN_u}$ is the block diagonal UL  frequency domain channel matrix between BS$_i$ and users in cell $i$. 
Accordingly, the DL  frequency domain channel matrix is its conjugate $\ubG_{j,i}^H$ in the TDD system.
Then the received signal at user $u$ in cell $i$ for subcarrier $k$ is given as
\begin{align}
    \label{eq:yiu_dl_ofdm}
    y_{i,u}^{\rm dl}(k) = \alpha\bg_{i,i,u}^H(k)\bw_{i,u}(k)s^{\rm dl}_{i,u}(k) + \alpha \!\!\! \sum_{(j,v) \neq (i,u)}^{N_c, N_u}\!\!\bg_{j,i,u}^H(k)\bw_{j,v}(k)s^{\rm dl}_{j,v}(k) + \tilde{q}^{\rm dl}_{i,u}(k) + \tilde{n}_{i,u}^{\rm dl}(k).
\end{align}

Based on \eqref{eq:Cqq_dl_ofdm} and \eqref{eq:yiu_dl_ofdm}, the DL SINR for user $u$ in cell $i$ at subcarrier $k$ is computed as 
\begin{align}
    \label{eq:sinr_dl_ofdm}
    &\Gamma^{\rm dl}_{i,u}(k) =\\
    \nonumber
    &\frac{\alpha^2|\bg_{i,i,u}^H(k)\bw_{i,u}(k)|^2}{\alpha^2 \sum_{(j,v) \neq (i,u)}^{N_c, N_u}\!|\bg_{j,i,u}^H(k)\bw_{j,v}(k)|^2 \!+\! \alpha(1\!-\!\alpha)\!\sum_{j=1}^{N_c}\ubg_{j,i,u}^H\!(k){\pmb \Psi}_{N_b}{\rm diag}\big({\pmb \Psi}_{N_b}^H\ubW_j\ubW_j^H{\pmb \Psi}_{N_b}\big){\pmb \Psi}_{N_b}^H\ubg_{j,i,u}\!(k)\! +\! 1}
\end{align}
where $\ubg_{j,i,u}\!(k)$ denotes the ($kN_u + u$)th column of $\ubG_{j,i}$, i.e., the entire column of $\ubG_{j,i}$ that corresponds to the channel  for $k$th subcarrier of user $u$.
Using \eqref{eq:sinr_dl_ofdm}, the DL OFDM transmit power minimization problem is formulated  as 
\begin{align}
    \label{eq:problem_dl_ofdm}
    \cP4: \qquad \min_{{\bf w}_{i,u}(k)}\   \alpha\!\!\sum_{i,u,k} \!\!{\bf w}_{i,u}^H(k){\bf w}_{i,u}(k)\qquad\text{s.t. }\  \Gamma^{\rm dl}_{i,u}(k) \geq \gamma_{i,u,k}, \quad \forall \, i,u,k.
\end{align}

\subsection{Joint Beamforming and Power Control for Wideband OFDM Systems}

Unlike the narrowband system, the quantization noise terms coupled with not only beamformers and transmit power but also OFDM modulation are the main challenge for showing the duality.
In the following theorem, we prove the duality by handling this issue.
\begin{theorem}[Duality]
    \label{thm:duality_ofdm}
    The duality holds between $\cP3$ and $\cP4$.
\begin{proof}
    Let ${\bf z}_{i,u}(k)$ be the interference-plus-noise term of \eqref{eq:yk_ul_ofdm} and $\bF_i(k)$ be the MMSE equalizer $\bF_i(k) =  {\bf C}_{{\bf z}_{i,u}(k){\bf z}_{i,u}(k)}^{-1} {\bf g}_{i,i,u}(k) $ where 
    \begin{align}
        \nonumber
        {\bf C}_{{\bf z}_{i,u}(k){\bf z}_{i,u}(k)} =\  &\alpha^2\!\!\! \sum_{(j,v)\neq (i,u)} \!\!\lambda_{j,v}(k) {\bf g}_{i,j,v}(k){\bf g}_{i,j,v}^H(k) + \alpha^2 {\bf I}_{N_b} \\
        \label{eq:Czz_ul_ofdm}
        &+ \alpha(1-\alpha) {\pmb \Psi}_{N_b}(k){\rm diag}\Big(\sum_{j=1}^{N_c}\underline{\bH}_{i,j}{\pmb \Psi}_{N_u}^H\underline{\pmb \Lambda}_j{\pmb \Psi}_{N_u}\underline{\bH}_{i,j}+\bI_{KN_b}\Big){\pmb \Psi}_{N_b}^H(k).
    \end{align}
    Noting that ${\pmb \Psi}_{N_b}^H\ubG_{i,j} = \ubH_{i,j}{\pmb \Psi}_{N_u} ^H$, we first rewrite the diagonal matrix in \eqref{eq:Czz_ul_ofdm} as
    \begin{align}
        \label{eq:diag_ul_ofdm}
        {\rm diag}\Big(\sum_{j=1}^{N_c}\underline{\bH}_{i,j}{\pmb \Psi}_{N_u}^H\underline{\pmb \Lambda}_j{\pmb \Psi}_{N_u}\underline{\bH}_{i,j}+\bI_{KN_b}\Big) = {\rm diag}\Big({\pmb \Psi}_{N_b}^H\ubG_i \,\underline{\pmb \Lambda}\,\ubG_i^H{\pmb \Psi}_{N_b}+\bI_{KN_b}\Big).
    \end{align}
    where $\ubG_i = [\ubG_{i,1},\dots, \ubG_{i,N_c}]$ and $\underline{\pmb \Lambda} = {\rm blkdiag}\big(\underline{\pmb \Lambda}_{1},\dots, \underline{\pmb \Lambda}_{N_c}\big)$.
    Following the same steps in the proof of Theorem~\ref{thm:duality} with \eqref{eq:diag_ul_ofdm} and ${\pmb \Psi}_{N_b}(k){\pmb \Psi}_{N_b}^H(k) = \bI_{N_b}$, $\cP3$ with the MMSE equalizer becomes
    \begin{gather}
        \label{eq:problem_ul_simple_ofdm}
        \min\  \sum_{i,u,k} \lambda_{i,u}(k)\\
        \nonumber
        \text{s.t. }\bar{\bf K}_{i,k}(\underline{\pmb \Lambda}) \preceq \alpha \bigg(1 + \frac{1}{\gamma_{i,u}(k)}\bigg)\lambda_{i,u}(k)  {\bf g}_{i,i,u}(k){\bf g}_{i,i,u}^H(k), \quad \forall i,u,k.
    \end{gather}
    where 
    \begin{align}
        \nonumber
        \bar{\bf K}_{i,k}(\underline{\pmb \Lambda}) = {\bf I}_{N_b} \! + \!\alpha \sum_{j,v} \lambda_{j,v}(k){\bf g}_{i,j,v}(k){\bf g}_{i,j,v}^H(k)\! +\! (1\!-\!\alpha){\pmb \Psi}_{N_b}(k){\rm diag}\Big({\pmb \Psi}_{N_b}^H\ubG_i \,\underline{\pmb \Lambda}\,\ubG_i^H{\pmb \Psi}_{N_b}\Big){\pmb \Psi}_{N_b}^H(k).
    \end{align}
   
    We need to show that \eqref{eq:problem_ul_simple_ofdm} is equivalent to the Lagrangian dual problem of $\cP4$.
    Similarly to the proof of Theorem~\ref{thm:duality}, the Lagrangian of $\cP4$ is given in the rearranged form as
    \begin{align}
        \nonumber
        &\bar{\cL} = \sum_{i,u,k}\mu_{i,u}(k) +\alpha(1\!-\!\alpha)\sum_{i,u,k}\mu_{i,u}(k)\sum_{j}\ubg_{j,i,u}^H\!(k){\pmb \Psi}_{N_b}{\rm diag}\big({\pmb \Psi}_{N_b}^H\ubW_j\ubW_j^H{\pmb \Psi}_{N_b}\big){\pmb \Psi}_{N_b}^H\ubg_{j,i,u}\!(k) \  +\\
        \label{eq:lagrangian_ofdm}
        &\sum_{i,u,k}\!\bw_{i,u}^H(k)\!\Bigg(\!\alpha\bI_{N_b}\! -\! \alpha^2\!\left(\!1\!+\!\frac{1}{\gamma_{i,u,k}}\!\right)\!\mu_{i,u}(k)\bg_{i,i,u}(k)\bg_{i,i,u}^H(k)\! +\! \alpha^2\!\sum_{j,v}\mu_{j,v}(k)\bg_{i,j,v}(k)\bg_{i,j,v}^H(k)\!\Bigg)\bw_{i,u}(k). 
    \end{align}
    We  rewrite the quantization error term in \eqref{eq:lagrangian_ofdm} to manipulate $\ubW_j$ in the diagonal matrix.
    Changing the indices of $\sum_{i,u,k}\mu_{i,u}(k)\sum_{j}\ubg_{j,i,u}^H\!(k){\pmb \Psi}_{N_b}{\rm diag}\big({\pmb \Psi}_{N_b}^H\ubW_j\ubW_j^H{\pmb \Psi}_{N_b}\big){\pmb \Psi}_{N_b}^H\ubg_{j,i,u}\!(k)$ from $(i,u,k,j)$ to $(j,v,\ell,i)$, we have
    \begin{align}
        \nonumber
        &\sum_{j,v,\ell,i}\mu_{j,v}(\ell)\ubg_{i,j,v}^H\!(\ell){\pmb \Psi}_{N_b}{\rm diag}\big({\pmb \Psi}_{N_b}^H\ubW_i\ubW_i^H{\pmb \Psi}_{N_b}\big){\pmb \Psi}_{N_b}^H\ubg_{i,j,v}\!(\ell)\\
        \label{eq:quant_noise_dl_ofdm1}
        &=\sum_{j,v,\ell,i}\mu_{j,v}(\ell)\ubg_{i,j,v}^H\!(\ell){\pmb \Psi}_{N_b}{\rm diag}\left(\sum_{u,k}|{\pmb \psi}_{N_b,m}^H(n)\ubw_{i,u}(k)|^2, \forall m,n\right){\pmb \Psi}_{N_b}^H\ubg_{i,j,v}\!(\ell)
    \end{align}
    where ${\pmb \psi}_{N_b,m}(n)$ denotes the $(m+(n-1)N_b)$th column of ${\pmb \Psi}_{N_b}$, i.e., ${\pmb \psi}_{N_b,m}(n) = [\bw_{{\rm DFT},n}\otimes\bI_{N_b}]_{:,m}$ for $m= 1,\dots,N_b$, $n=1,\dots,K$,   and $\ubw_{i,u}(k)$ is the $(kN_u+u)$th column of $\ubW_i$, i.e., the entire column of $\ubW_i$ that corresponds to the precoder for $k$th subcarrier of user $u$.
    Let  $\bM_i(k) = {\rm diag}(\mu_{i,1}(k),\dots,\mu_{i,N_u}(k))$, $\ubM_i = {\rm blkdiag}\big(\bM_i(0),\dots,\bM_i(K-1)\big)$, and $\ubM = [\ubM_{1},\dots, \ubM_{N_c}]$.
    Recalling that ${\pmb \Psi}_{N_b}(k) = \big([\bW_{{\rm DFT}}]_{k+1,:}\otimes \bI_{N_b}\big)$ and $\ubG_i = [\ubG_{i,1},\dots, \ubG_{i,N_c}]$,  \eqref{eq:quant_noise_dl_ofdm1} is rewritten as
    \begin{align}
        \nonumber
        &\sum_{j,v,\ell,i}\mu_{j,v}(\ell)\sum_{m,n}\left(\sum_{u,k}|{\pmb \psi}_{N_b,m}^H(n)\ubw_{i,u}(k)|^2\left(\sum_{r}\underline{g}^*_{i,j,v,r}(\ell)\psi_{N_b,m,r}(n)\right)\left(\sum_{r'}\underline{g}_{i,j,v,r'}(\ell)\psi^*_{N_b,m,r'}(n)\right)\right)\\
        \nonumber
        & = \sum_{i,u,k}\ubw_{i,u}^H(k)\left(\sum_{m,n}{\pmb \psi}_{N_b,m}(n)\left(\sum_{j,v,\ell}\mu_{j,v}(\ell){\pmb \psi}_{N_b,m}^H(n)\ubg_{i,j,v}(\ell){\ubg}^H_{i,j,v}(\ell){\pmb \psi}_{N_b,m}(n)\right)\!{\pmb \psi}_{N_b,m}^H(n)\!\right)\ubw_{i,u}(k)\\
        \nonumber
        & = \sum_{i,u,k}\ubw_{i,u}^H(k)\left(\sum_{m,n}{\pmb \psi}_{N_b,m}(n){\pmb \psi}_{N_b,m}^H(n)\ubG_i\,\ubM\,\ubG_i^H{\pmb \psi}_{N_b,m}(n){\pmb \psi}_{N_b,m}^H(n)\right)\ubw_{i,u}(k)\\
        \nonumber
        & = \sum_{i,u,k}\ubw_{i,u}^H(k){\pmb \Psi}_{N_b}{\rm diag}\left({\pmb \Psi}_{N_b}^H\ubG_i\,\ubM\,\ubG_i^H{\pmb \Psi}_{N_b}\right){\pmb \Psi}_{N_b}^H\ubw_{i,u}(k)\\
        \label{eq:quant_noise_dl_ofdm2}
        & \stackrel{(a)}= \sum_{i,u,k}\bw_{i,u}^H(k){\pmb \Psi}_{N_b}(k){\rm diag}\left({\pmb \Psi}_{N_b}^H\ubG_i\,\ubM\,\ubG_i^H{\pmb \Psi}_{N_b}\right){\pmb \Psi}_{N_b}^H(k)\bw_{i,u}(k).
    \end{align}
   Here $(a)$ comes from $\ubw_{i,u}^H(k){\pmb \Psi}_{N_b} = \bw_{i,u}^H(k){\pmb \Psi}_{N_b}(k)$ as $\ubw_{i,u}(k)$ has nonzero elements $\bw_{i,u}(k)$ only in the place that corresponds to the precoder for subcarrier $k$, and $\underline{g}_{i,j,v,r}(\ell)$ and ${\psi}_{N_b,m,r}(n)$ are the $r$th elements of $\ubg_{i,j,v}(\ell)$ and ${\pmb \psi}_{N_b,m}(n)$, respectively.
    
    Applying \eqref{eq:quant_noise_dl_ofdm2} to the Lagrangian in \eqref{eq:lagrangian_ofdm}, we have
    \begin{align}
        \nonumber
        &\bar{\cL} = \sum_{i,u,k}\mu_{i,u}(k) +\sum_{i,u,k}\!\bw_{i,u}^H(k)\!\Bigg(\!\alpha\bI_{N_b}\! -\! \alpha^2\!\left(\!1\!+\!\frac{1}{\gamma_{i,u,k}}\!\right)\!\mu_{i,u}(k)\bg_{i,i,u}(k)\bg_{i,i,u}^H(k) \ +\\
        &\alpha^2\!\sum_{j,v}\mu_{j,v}(k)\bg_{i,j,v}(k)\bg_{i,j,v}^H(k)\!+\!\alpha(1\!-\!\alpha){\pmb \Psi}_{N_b}(k){\rm diag}\!\left({\pmb \Psi}_{N_b}^H\ubG_i\,\ubM\,\ubG_i^H{\pmb \Psi}_{N_b}\!\right)\!{\pmb \Psi}_{N_b}^H(k)\!\Bigg)\bw_{i,u}(k).
    \end{align}
    Following similar steps in the proof of Theorem~\ref{thm:duality}, the Lagrangian dual problem of $\cP4$ becomes
    \begin{gather}
        \label{eq:dual_problem_ofdm}
        \max_{\mu_{i,u}} \sum_{i,u,k} \mu_{i,u}(k) \\
        \nonumber
        {\text{s.t. }}\ \bar{\bf K}_{i,k}(\ubM) \succeq \alpha \bigg(1 + \frac{1}{\gamma_{i,u,k}}\bigg)\mu_{i,u}(k)  {\bg}_{i,i,u}(k){\bg}_{i,i,u}^H(k)  , \quad \forall \, i,u,k
    \end{gather}
    where  
    \begin{align}
        \nonumber
        \bar{\bf K}_{i,k}(\ubM)\! =\! \bI_{N_b} \!+\!
        \alpha\!\sum_{j,v}\mu_{j,v}(k)\bg_{i,j,v}(k)\bg_{i,j,v}^H(k)\!+\!(1\!-\!\alpha){\pmb \Psi}_{N_b}(k){\rm diag}\!\left({\pmb \Psi}_{N_b}^H\ubG_i\,\ubM\,\ubG_i^H{\pmb \Psi}_{N_b}\!\right)\!{\pmb \Psi}_{N_b}^H(k).
    \end{align}
    Since the problem in \eqref{eq:dual_problem_ofdm} has its optimal solution when the constraints are active, it is also equivalent to \eqref{eq:problem_ul_simple_ofdm}. 
    This completes the proof.
\end{proof}
\end{theorem}

\begin{corollary}[Strong duality]
    \label{cor:strong_duality_ofdm}
    Strong duality holds for $\cP4$ and its Lagrangian dual problem. 
    \begin{proof}
        We use \eqref{eq:quant_noise_dl_ofdm2} to manipulate the precoders $\bW_i(k)$ in the diagonal matrix of the quantization term in the SINR \eqref{eq:sinr_dl_ofdm}, and follow similar approach as the proof of Corollary~\ref{cor:strong_duality}.
        Then $\cP4$ can be cast to the SOCP. 
        In addition, $\cP4$ is strictly feasible. 
        This completes the proof.
    \end{proof}
\end{corollary}
Since we have shown that the duality between $\cP3$ and $\cP4$ with no duality gap, we can characterize the optimal solutions via the duality. 
Here we briefly describe the overall procedures as they are similar to the narrowband case;
solving Karush-Kuhn-Tucker (KKT) conditions, we can show that the UL ODFM problem $\cP3$ can be solved by the distributed iterative algorithm that is proposed in Sec.~\ref{subsec:algorithm} with the following solution:
\begin{align}
    \label{eq:solution_ofdm}
    \lambda_{i,u}(k) = \frac{1}{\alpha \Big(1+\frac{1}{\gamma_{i,u,k}}\Big){\bf g}_{i,i,u}^H(k) \bar{\bf K}_{i,k}^{-1}(\underline{\pmb\Lambda}) {\bf g}_{i,i,u}(k)}.
\end{align}
Note that \eqref{eq:solution_ofdm} needs to be computed over not only users but also subcarriers at each BS.
Now let $\lambda_{i,u}^{(n+1)} = f_{i,u,k}\big(\underline{\pmb \Lambda}^{(n)}\big)$. 
Then, as in the proof of Corollary~\ref{cor:convergence}, the convergence of the iterative method can be proved by showing that $f_{i,u,k}\big(\underline{\pmb \Lambda}^{(n)}\big)$ is a standard function.
Using the obtained optimal UL transmit power $\lambda_{i,u}(k)$ from the standard fixed-point iteration, the MMSE equalizer $\bF_i(k)$ for the received signal at each subcarrier $\by_i^{\rm ul}(k)$ is computed as $\bF_i(k) =  {\bf C}_{{\bf z}_{i,u}(k){\bf z}_{i,u}(k)}^{-1} {\bf g}_{i,i,u}(k)$ where $ {\bf C}_{{\bf z}_{i,u}(k){\bf z}_{i,u}(k)}^{-1} $ is given in \eqref{eq:Czz_ul_ofdm}.
Based on $\bF_i(k)$, we can also obtain the optimal precoder $\bW_i(k)$ for $\cP4$ from appropriate scaling of $\bF_i(k)$ as shown in Corollary~\ref{cor:dl_precoder_ofdm}.

\begin{corollary}[Precoder]
    \label{cor:dl_precoder_ofdm}
    With the proper scaling coefficient in wideband case, an optimal DL precoder can be proportional to the uplink MMSE receiver, i.e.,  $\bw_{i,u}(k) = \sqrt{\underline{\tau}_{i,u}(k)}\bff_{i,u}(k) \;\forall i, u, k$, and $\underline{\tau}_{i,u}$ is derived as $\underline{\btau} = \underline{\bSigma}^{-1}{\bf 1}$, where  ${\bf 1}$ is a $N_uN_cK \times 1$ column vector, $\underline{\btau}=[\underline{\btau}^T(0), \cdots, \underline{\btau}^T(K-1)]^T$ with $\underline{\btau}(k) = [\underline{\btau}_{1}^T(k), \underline{\btau}_{2}^T(k), \cdots, \underline{\btau}_{N_c}^T(k) ]^T$ and  $\underline{\btau}_{i}^T(k) = [\underline{\tau}_{i,1}(k), \underline{\tau}_{i,2}(k), \cdots, \underline{\tau}_{i,N_u}(k)]^T$, and $\underline{\bSigma} = {\rm blkdiag} (\underline{\bSigma}(0),\dots,\underline{\bSigma}(K-1))$ whose submatrix is composed as
    \begin{equation}
    \label{eq:wb_constraint_matrix)}
    \underline{\bSigma}(k) = 
    \begin{pmatrix}
    \underline{\bSigma}_{1,1}(k) & \underline{\bSigma}_{1,2}(k) & \cdots & \underline{\bSigma}_{1,N_c}(k) \\
    \underline{\bSigma}_{2,1}(k) & \underline{\bSigma}_{2,2}(k) & \cdots & \underline{\bSigma}_{2,N_c}(k) \\
    \vdots  & \vdots  & \ddots & \vdots  \\
    \underline{\bSigma}_{N_c,1}(k) & \underline{\bSigma}_{N_c,2}(k) & \cdots & \underline{\bSigma}_{N_c,N_c}(k)
     \end{pmatrix},
    \end{equation}
    and
    \begin{align} 
        &[\underline{\bSigma}_{i,j}(k)]_{u,v} \nonumber \\
        &=\begin{cases}
          \frac{\alpha^2}{\gamma_{i,u}(k)}|\bg_{i,i,u}^H(k)\bff_{i,u}(k)|^2 \! \\
         \quad - \alpha(1\!-\!\alpha)\!\sum_{n}\bff_{i,u}^H(n){\pmb \Psi}_{N_b}(n){\rm diag}\big({\pmb \Psi}_{N_b}^H\ubg_{i,i,u}\!(k)\ubg_{i,i,u}^H\!(k){\pmb \Psi}_{N_b}\big){\pmb \Psi}_{N_b}^H(n)\bff_{i,u}(n)  \quad\!\! \text{if } i=j \text{, } u=v, \\
            \nonumber
        - \alpha^2 |\bg_{j,i,u}^H(k)\bff_{j,v}(k)|^2\! \\
       \quad - \alpha(1\!-\!\alpha)\!\sum_{n}\bff_{j,v}^H(n){\pmb \Psi}_{N_b}(n){\rm diag}\big({\pmb \Psi}_{N_b}^H\ubg_{j,i,u}\!(k)\ubg_{j,i,u}^H\!(k){\pmb \Psi}_{N_b}\big){\pmb \Psi}_{N_b}^H(n)\bff_{j,v}(n) \quad \text{otherwise.}
        \end{cases}
    \end{align}
    
    \begin{proof}
        See Appendix \ref{appx:DL_precoder_ofdm}.
    \end{proof}
\end{corollary}
Therefore, since the strong duality also holds between the UL and DL wideband OFDM systems with low-resolution ADCs and low-resolution DACs, respectively, we have shown that $\cP4$ can also be solved by using the distributed iterative algorithm as the narrowaband case.

\section{Simulation Results}

In this section, we validate the derived theoretical results and  the proposed quantization-aware iterative CoMP (Q-iCoMP) algorithm and deterministic CoMP (Q-dCoMP) algorithm. 
We also simulate the quantization-aware per-cell (Q-Percell) based iterative algorithm by adapting the per-cell algorithm in \cite{rashid1998transmit} to low-resolution ADC systems based on the derived SINR with quantization noise in \eqref{eq:sinr_ul}. 
For the Q-Percell algorithm, each BS first finds its optimal solution based on the iterative algorithm in \cite{rashid1998transmit} by considering the inter-cell interference as noise and assuming it to be fixed. 
Once the BSs derive solutions for the given noise power, they update the noise power and find solutions again.
These steps are repeated until the solutions converge.
For simulations, we use networks with $N_c \in \{2,7\}$ cells.
For $N_c=2$, two cells are adjacent to each other. 
For $N_c=7$, six cells are adjacent to a center cell.  
Assuming narrowband communications, we consider each BS to be located in the center of each hexagonal cell and randomly distribute $N_u$ users in each cell.
For small scale fading, we assume Rayleigh channels with a zero mean and unit variance. 
For large scale fading, we adopt the log-distance pathloss model in \cite{erceg1999empirically}. 
The distance between adjacent BSs is $2\  \rm km$ and the minimum distance between  BSs and users is $100\  \rm m$.
Considering a $2.4 \ \rm GHz$ carrier frequency with $10\  \rm MHz$ bandwidth, we use $8.7\  \rm dB$ lognormal shadowing variance and $5 \ \rm dB$ noise figure.
For simplicity, we assume that the target SINR $\gamma$ is equal for all users across all cells.


\begin{figure}[!t]\centering
\includegraphics[scale = 0.45]{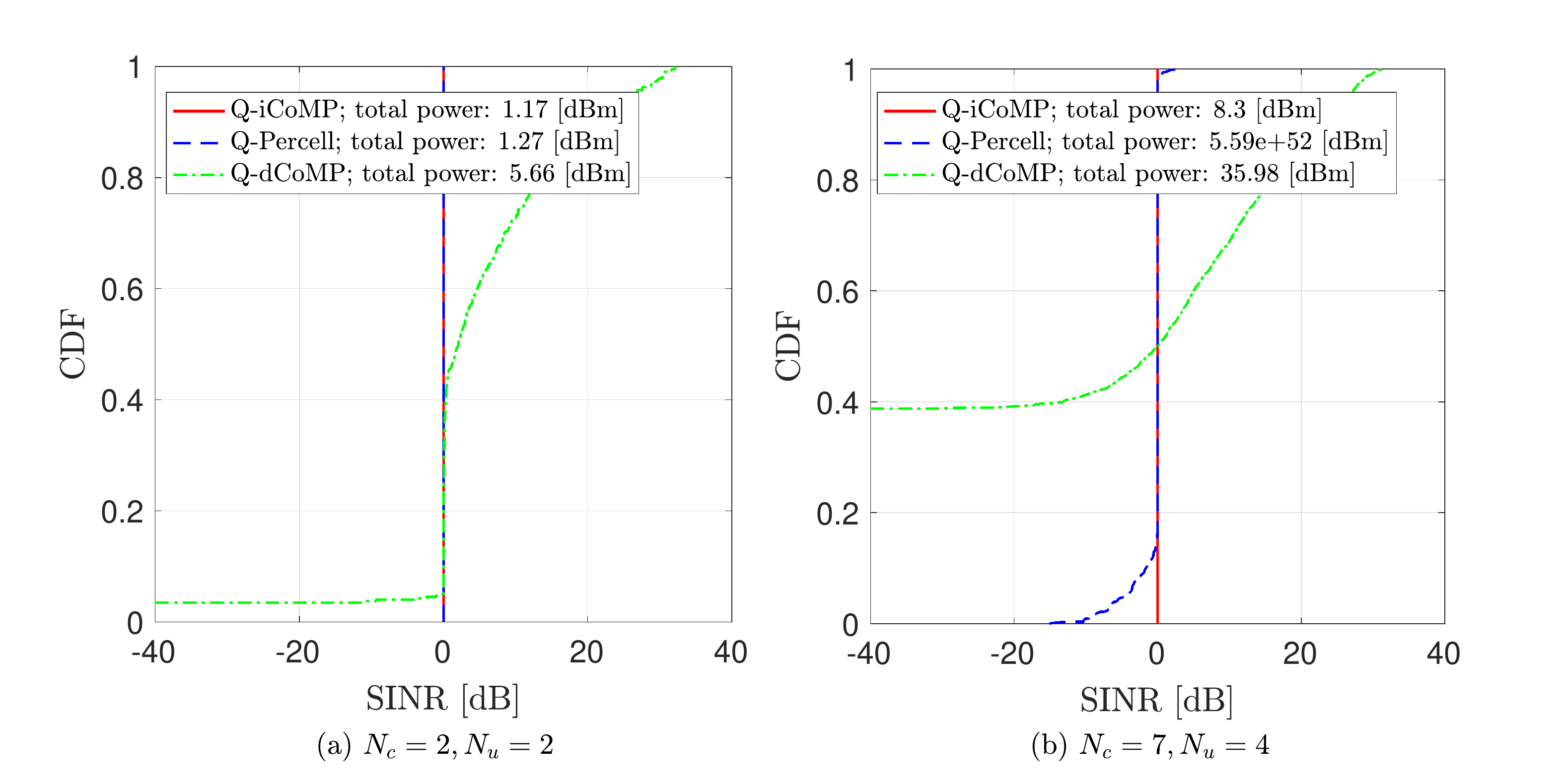}
\vspace{-1.5em}
\caption{CDFs of the SINRs of users in all cells for $\gamma = 0\  \rm dB$ target SINR, $b= 3$ quantiation bits, and $N_b = 64$ BS antennas with (a) $N_c = 2$ cells with $N_u = 2$ users per cell and with (b) $N_c = 7$ cells with $N_u = 4$ users per cell.} 
\label{fig:cdf}
\end{figure}

Fig.~\ref{fig:cdf} shows the cumulative density function (CDF) of the SINR of users in all cells for  $\gamma = 0\  \rm dB$, $b=3$, and $N_b = 64$ with (a) $(N_c = 2 ,N_u = 2)$ and with (b) $(N_c = 7, N_u = 4)$.
The proposed Q-iCoMP algorithm shows a step function-like CDF at $0\ \rm dB$ SINR with the minimum total transmit power among the evaluated algorithms for both cases (a) and (b). 
This validates the performance of the Q-iCoMP algorithm which provides an optimal solution for the UL and DL problems in \eqref{eq:problem_ul} and \eqref{eq:problem_dl}.
Although the Q-Percell algorithm achieves similar SINR results with slightly increased total transmit power for (a) ($N_c = 2, N_u = 2$), about $10\%$ of users have the SINR less than the target SINR and the total transmit power becomes excessive for  (b) ($N_c = 7, N_u = 4$).
Accordingly, the Q-Percell algorithm is only feasible when the numbers of cooperating BSs and associated users are small.
Regarding the deterministic approach, more than $95\%$ of users meets the target SINR  for (a)  ($N_c = 2, N_u = 2$).
For (b) ($N_c = 7, N_u = 4$), however,  about $50\%$ of users cannot achieve the target SINR, and most of them have zero transmit power.
Although the Q-Percell algorithm shows better performance in satisfying the target SINR than the Q-dCoMP algorithm, its total transmit power can easily diverge when the network becomes denser. 
Therefore, the Q-iCoMP algorithm achieves the best performance and the Q-dCoMP algorithm can be more practical than the Q-Percell algorithm for dense networks.

\begin{figure}[!t]\centering
\includegraphics[scale = 0.55]{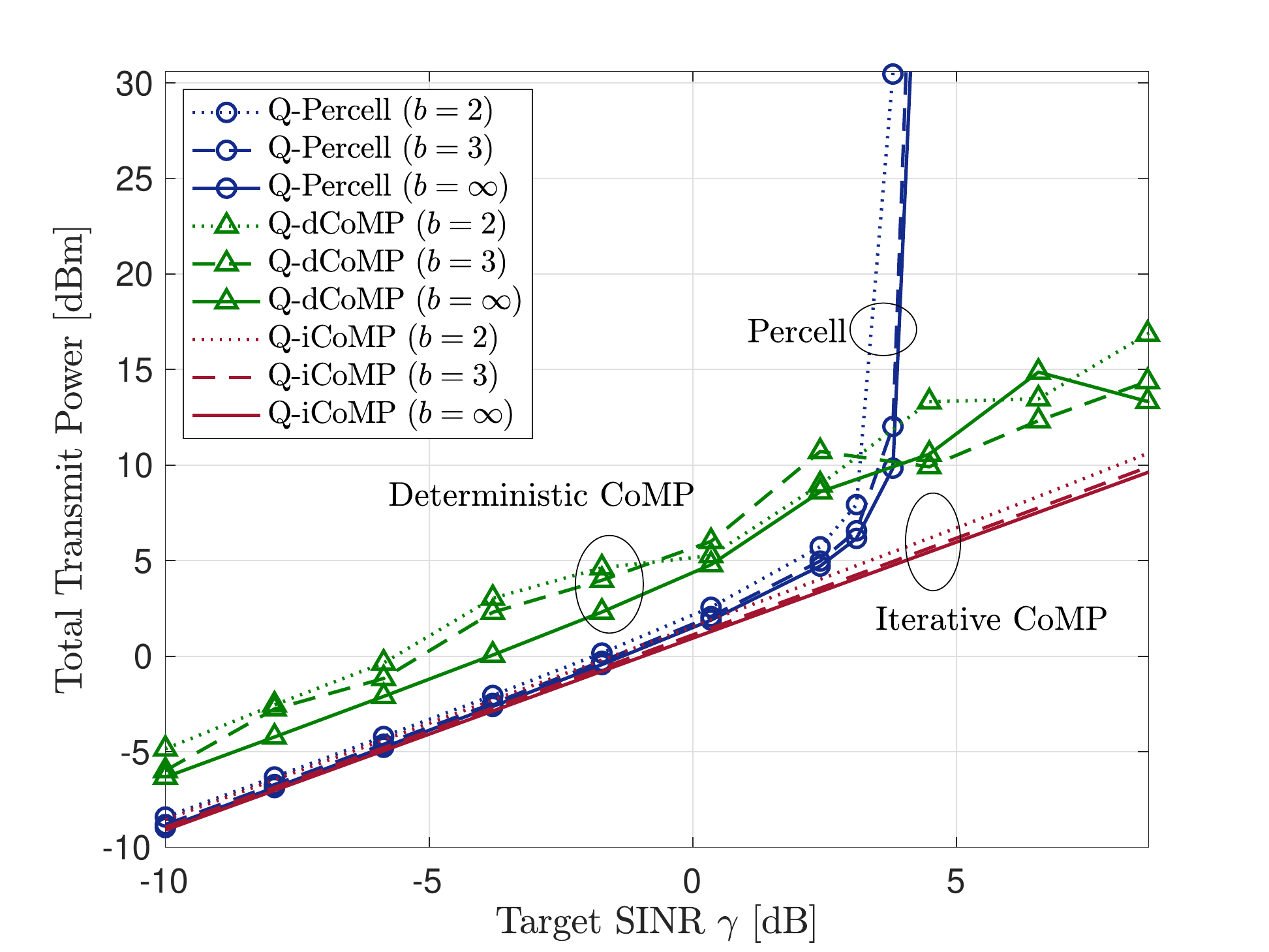}
\vspace{-1em}
\caption{Total transmit power versus the target SINR for $N_b = 64$ BS antennas, $N_c = 2$ cells and $N_u = 2$ users per cell.} 
\label{fig:txpower}
\end{figure}

Fig.~\ref{fig:txpower} shows the total transmit power versus the target SINRs for $N_b = 64$, $N_c = 2$, $N_u = 2$, and $b \in \{2, 3, \infty\}$.
For the considered target SINR range, the Q-iCoMP algorithm shows the minimum total transmit power.
The increase in the transmit power due to the increased quantization error is also small.
Despite that the Q-Percell algorithm also achieves similar performance at the low to medium target SINR, the transmit power of the algorithm diverges in the medium to high target SINR range. 
The Q-dCoMP algorithm shows a larger gap between different quantization resolutions than that in the iterative algorithms.
We note that as the target SINR increases, the total transmit power curves of the Q-dCoMP algorithm show larger fluctuation, and there are crossing points between different resolutions; as the target SINR increases, more BSs are likely to assign with zero transmit power to reduce interference to the other cells, which happens more often with a less number of quantization bits.

\begin{figure}[!t]\centering
\includegraphics[scale = 0.4]{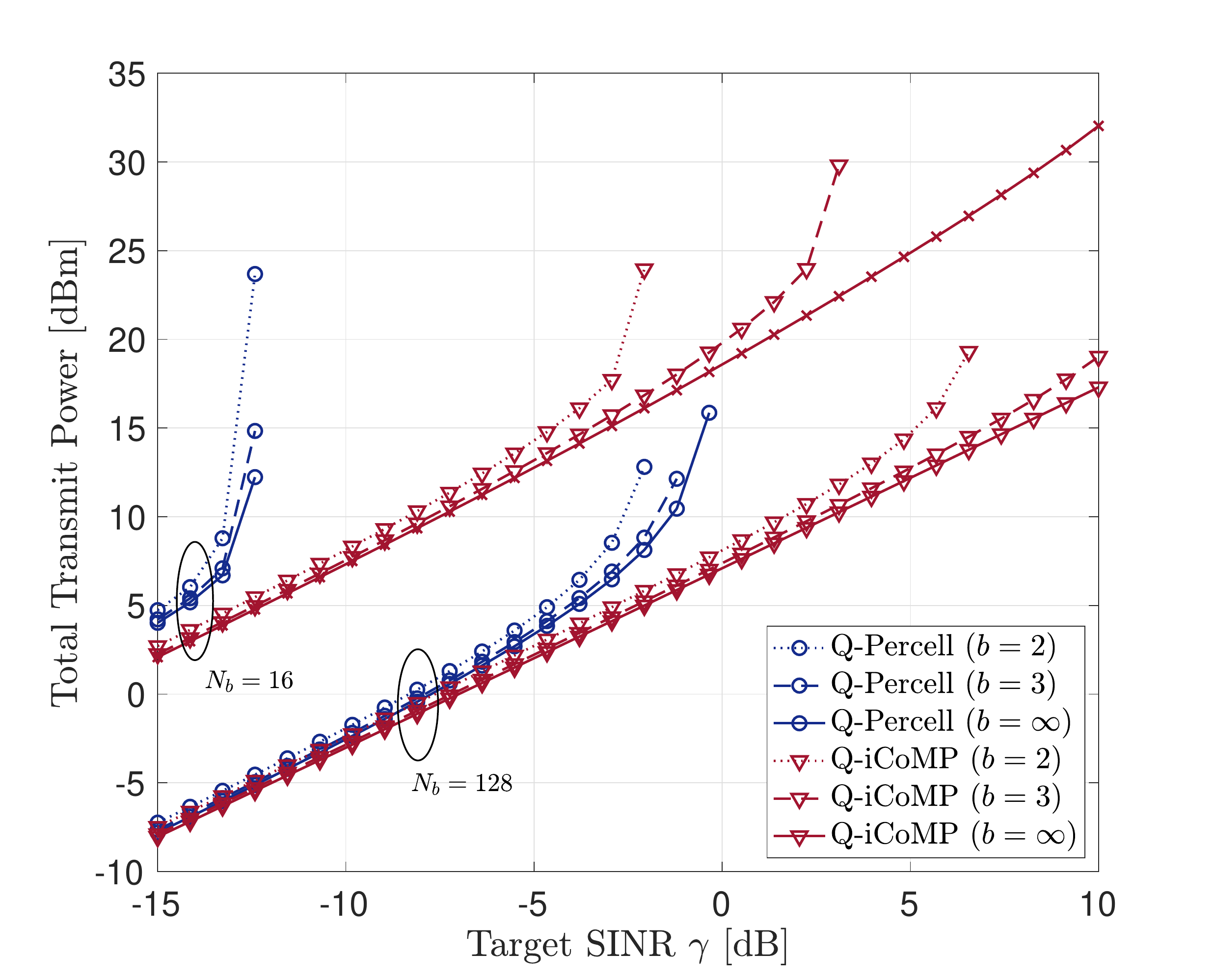}
\vspace{-1em}
\caption{Total transmit power versus the target SINR for $N_b \in \{16, 128\}$, $N_c = 7$ cells, and $N_u = 4$ users per cell.} 
\label{fig:txpower2}
\end{figure}

In Fig.~\ref{fig:txpower2}, the network with $N_c = 7$ and $N_u = 4$ is considered for the different $b$ and $N_b$.
For $N_b=16$, the Q-Percell algorithm is almost infeasible and the Q-iCoMP algorithm also shows divergence in the total transmit power at the medium to high target SINRs with a small number of quantization bits.
Increasing the number of BS antennas from $16$ to $128$ provides more than $10$ dB SINR gain.
Accordingly, for $N_b =128$ which is considered as the massive MIMO system, the Q-iCoMP algorithm achieves the target SINRs for all users without divergence even with $b = 3$, whereas the Q-Percell algorithm still suffers from excessive power consumption in the medium to high target SINR range.
Therefore, in massive MIMO systems, the coordinated joint BF and PC can provide reliable and power-efficient communications even with a small number of quantization bits, thereby achieving spectrum- and energy-efficient communications.

\section{Conclusion}

This paper investigated coordinated multipoint beamforming and power control for massive MIMO systems with low-resolution ADCs and DACs.
We showed that strong duality holds between UL and DL total transmit power minimization problems under target SINR constraints in low-resolution ADC and DAC systems based on the additive quantization noise model.
Leveraging the duality, a fixed-point CoMP algorithm was proposed to jointly solve the UL and DL problems by incorporating the coarse quantization effect.
The proposed algorithm provides optimal solutions for the UL and DL problems in an efficient and distributed manner without requiring explicit out-of-cell channel estimation. 
In addition, a deterministic algorithm was developed to provide a closed-form solution for the UL problem with the assumption of homogeneous transmit powers and SINR constraints within each cell.
We proved that the derived results can be extended to wideband OFDM systems when optimizing a beamformer and transmit power for each user and subcarrier under coarse quantizaiton. 
Via simulations, we showed that the proposed iterative CoMP algorithm can achieve high target SINRs without divergence of transmit power for low-resolution ADC and DAC systems, whereas the conventional per-cell based algorithm suffers from excessive power consumption even with infinite-resolution ADCs and DACs. 
We also observe that the deterministic solution can achieve a reasonable trade-off between total transmit power and achieved SINR.
Overall, in massive MIMO systems integrated with coarse quantization, the coordinated beamforming and power control offers spectrum- and power-efficient wireless communication systems.

\begin{appendices}

\section{Proof of Corollary~\ref{cor:strong_duality}}
\label{appx:strong_duality}

We first show that \eqref{eq:problem_dl} can be represented as a standard conic optimization problem.  
    Let ${\bf W}$ be defined as ${\bf W} = [{\bf W}_1, \cdots, {\bf W}_{N_c}]$, then the DL problem \eqref{eq:problem_dl} is rewritten as
    \begin{gather}
        \label{eq:strong_pf}
        \min_{{\bW},P_o} P_o \\ 
        \label{eq:strong_pf1}
        {\rm s.t.}\ \Gamma_{i,u}^{\rm dl} \geq \gamma_{i,u}, \quad \forall i,u\\ 
        \label{eq:strong_pf2}
        {\rm Tr}\big({\bf W}^H{\bf W}\big) \leq P_o
    \end{gather}
    where $P_o$ is a positive slack variable.
    As in \cite{wiesel2005linear, yu2007transmitter}, we can take a diagonal phase scaling on the right of each precoder as ${\bf W}_i{\rm diag}(e^{j\phi_{i,1}},\dots,e^{j\phi_{i,{N_u}}})$ for $i = 1,\cdots, N_c$, without changing the objective nor the constraints, we can design the precoder to be ${\bf w}_{i,u}^H{\bf h}_{i,i,u} \geq 0$, $\forall i,u$ .
    
    Using \eqref{eq:lagrangian_pf2}, we rewrite the quantization term in \eqref{eq:sinr_dl} as
    \begin{align}
        \nonumber
        \sum_j\bh_{j,i,u}^H\bC_{\bq^{\rm dl}_j\bq^{\rm dl}_j}\bh_{j,i,u} &= \alpha(1-\alpha) \sum_j\bh_{j,i,u}^H{\rm diag}(\bW_j\bW_j^H)\bh_{j,i,u}\\
        \label{eq:strong_pf3}
        & = \alpha(1-\alpha)\sum_{j,v}\bw_{j,v}^H{\rm diag}(\bh_{j,i,u}\bh_{j,i,u}^H)\bw_{j,v}.
    \end{align}
    Let $\bD_{j,i,u} = {\rm diag}(\bh_{j,i,u}\bh_{j,i,u}^H)$, $\bW_{\rm BD} = {\rm blkdiag}(\bW_1,\dots,\bW_{N_c})$, and $\tilde{\bW}_{\rm BD} = {\rm blkdiag}((\bI_{N_b}\otimes\bW_1),\dots,(\bI_{N_b}\otimes\bW_{N_c}))$. 
    Using \eqref{eq:strong_pf3}, the SINR constraints in \eqref{eq:strong_pf1} can be rearranged as
    \begin{align}
        \label{eq:strong_pf4}
        \alpha^2 \bigg(1+\frac{1}{\gamma_{i,u}}\bigg) |{\bf w}_{i,u}^H {\bf h}_{i,i,u}|^2 \geq 
        \left\|
        \begin{matrix}
            \alpha\bW_{\rm BD}^H{\rm vec}(\bh_{1,i,u},\dots,\bh_{N_c,i,u}) \\
            \sqrt{\alpha(1-\alpha)}\tilde{\bW}_{\rm BD}^H{\rm vec}(\bD^{1/2}_{1,i,u},\dots,\bD^{1/2}_{N_c,i,u})\\
            1
        \end{matrix}
        \right\|^2, \quad \forall \, i,u.
    \end{align}
    Since we restrict the precoders to be ${\bf w}_{i,u}^H {\bf h}_{i,u} \geq 0$, we can take square root for  \eqref{eq:strong_pf4}.
    In addition, the power constraint \eqref{eq:strong_pf2} can be reformulated as $ \|{\rm vec}({\bf W})\| \leq \sqrt{P_o}$. 
    Using \eqref{eq:strong_pf4} and $ \|{\rm vec}({\bf W})\| \leq \sqrt{P_o}$, the problem in \eqref{eq:strong_pf} can be cast to the standard second order conic problem (SOCP) \cite{wiesel2005linear}.
    
    Next, \eqref{eq:problem_dl} is strictly feasible because given a solution ${\bf W}$, it can be always scaled by a factor of $c >1$ satisfying the constraints. 
    Thus, the strong duality holds between \eqref{eq:problem_ul} and \eqref{eq:problem_dl}. 
    \qed

\section{Proof of Corollary~\ref{cor:DL_precoder}}
\label{appx:DL_precoder}

    To find the optimal $\bw_{i,u}$, we set the derivative of the Lagrangian with respect to $\bw_{i,u}$ in \eqref{eq:zero_derivative} to zero, and solve it for $\bw_{i,u}$. 
    Then we have 
    \begin{align}
        \nonumber
        \bw_{i,u} &=  \left( \!\alpha^2\!\!\!\!\!\sum_{(j,v)\neq (i,u)} \!\!\!\lambda_{j,v} {\bf h}_{i,j,v}{\bf h}_{i,j,v}^H\! +\! \alpha {\bf I}_{N_b}\! +\! \alpha(1\!-\!\alpha){\rm diag}({\bf H}_i {\pmb \Lambda} {\bf H}_i^H)\!\right)^{-1}\!\!\!\alpha^2\!\left(\!1\!+\!\frac{1}{\gamma_{i,u}}\right)\!\lambda_{i,u}{\bf h}_{i,i,u}\bh_{i,i,u}^H\bw_{i,u}\\
        \nonumber
        & = \alpha^2\left(1+\frac{1}{\gamma_{i,u}}\right)\lambda_{i,u}\bh_{i,i,u}^H\bw_{i,u} \bff_{i,u}
    \end{align}
    where $\bff_{i,u}$ is in \eqref{eq:MMSE2}.
    We consider $\sqrt{\tau_{i,u}} =  \alpha^2\left(1+\frac{1}{\gamma_{i,u}}\right)\lambda_{i,u}\bh_{i,i,u}^H\bw_{i,u}$ and thus, $\bw_{i,u} = \sqrt{\tau_{i,u}}\bff_{i,u}$.
    
    Based on the Lagrangian dual problem, the global optimum occurs when the constraints satisfy equality conditions, i.e., active constraints.
    By replacing $\bw_{i,u}$ in \eqref{eq:sinr_dl} with $\sqrt{\tau_{i,u}}\bff_{i,u}$, the constraints of the primal DL problem satisfy the following conditions: 
    \begin{align}
        \nonumber
        &\frac{\alpha^2}{\gamma_{i,u}} |{\bf w}_{i,u}^H {\bf h}_{i,i,u}|^2 -  \alpha^2 \sum_{v \neq u} {| {\bf w}_{i,v}^H {\bf h}_{i,i,u}|^2} - \alpha^2 \sum_{\substack{j \neq i\\v}}|{\bf w}_{j,v}^H {\bf h}_{j,i,u}|^2  - \sum_{j}{\bf h}_{j,i,u}^H {\bf C}_{\bq^{\rm dl}_j\bq^{\rm dl}_j} {\bf h}_{j,i,u}  \\
        \nonumber
        &\stackrel{(a)}=  \frac{\alpha^2}{\gamma_{i,u}} |\bff_{i,u}^H {\bf h}_{i,i,u}|^2 \tau_{i,u} -  \alpha^2 \!\!\!\sum_{(j,v) \neq (i,u)}\!\! {| \bff_{j,v}^H {\bf h}_{j,i,u}|^2} \tau_{j,v} 
        - \alpha(1-\alpha)\sum_{j,v}\bff_{j,v}^H{\rm diag}(\bh_{j,i,u}\bh_{j,i,u}^H)\bff_{j,v}\tau_{j,v} \\
      \label{eq:dl_active}
        &= 1, \quad \forall i, u,
    \end{align} 
    where $(a)$ is from \eqref{eq:strong_pf3} and $\bw_{i,u} = \sqrt{\tau_{i,u}}\bff_{i,u}$. 
    We express \eqref{eq:dl_active} for all $i,u$ as a matrix form: $\bSigma\btau = {\bf 1}$.
    Therefore, $\tau_{i,u}$ can be obtained as $\btau = \bSigma^{-1}{\bf 1}$.
\section{Proof of Corollary~\ref{cor:dl_precoder_ofdm}}
\label{appx:DL_precoder_ofdm}

To guarantee the stationarity of the KKT condition with the DL constraint in \eqref{eq:problem_dl_ofdm}, the SINR of subcarrier $k$ of  user $u$ in cell $i$ needs to fulfill the target SINR with equality. 
To represent the SINR constraint in a tractable form, we rewrite the quantization error term in \eqref{eq:sinr_dl_ofdm}.
To this end, let us define $\mu_{i',u'}(n)$ where $\mu_{i',u'}(n)=1$ if $i'=i$, $u'=u$ and $n=k$, and $\mu_{i',u'}(n)=0$ otherwise. 
Then the quantization error term becomes
        \begin{align}
            \nonumber
            &\sum_{j=1}^{N_c}\ubg_{j,i,u}^H\!(k){\pmb \Psi}_{N_b}{\rm diag}\big({\pmb \Psi}_{N_b}^H\ubW_j\ubW_j^H{\pmb \Psi}_{N_b}\big){\pmb \Psi}_{N_b}^H\ubg_{j,i,u}\!(k) \\
            \nonumber
            &=\sum_{i',u',n,j}\mu_{i',u'}(n)\ubg_{j,i',u'}^H\!(n){\pmb \Psi}_{N_b}{\rm diag}\big({\pmb \Psi}_{N_b}^H\ubW_j\ubW_j^H{\pmb \Psi}_{N_b}\big){\pmb \Psi}_{N_b}^H\ubg_{j,i',u'}\!(n)\\
            \nonumber
            &\stackrel{(a)}=\sum_{j,v,\ell}\bw_{j,v}^H(\ell){\pmb \Psi}_{N_b}(\ell){\rm diag}\left({\pmb \Psi}_{N_b}^H\ubG_j\,\ubM\,\ubG_j^H{\pmb \Psi}_{N_b}\right){\pmb \Psi}_{N_b}^H(\ell)\bw_{j,v}(\ell)\\
            \label{eq:QN_reform_ofdm}
            &\stackrel{(b)}= \sum_{j,v,\ell}\bw_{j,v}^H(\ell){\pmb \Psi}_{N_b}(\ell){\rm diag}\left({\pmb \Psi}_{N_b}^H\ubg_{j,i,u}\!(k)\ubg_{j,i,u}^H\!(k)^H{\pmb \Psi}_{N_b}\right){\pmb \Psi}_{N_b}^H(\ell)\bw_{j,v}^H(\ell),
        \end{align}
        where $(a)$ comes from following the same steps in \eqref{eq:quant_noise_dl_ofdm1} and \eqref{eq:quant_noise_dl_ofdm2}. 
        Recalling the definition of $\ubM$ defined in the proof of Theorem \ref{thm:duality_ofdm} with slight abuse of notations, $(b)$ follows from  $\ubG_j\,\ubM\,\ubG_j^H = \ubg_{j,i,u}\!(k)\ubg_{j,i,u}^H(k)$.
        Replacing $\bw_{i,u}(n)$ with $\sqrt{\underline{\tau}_{i,u}(n)}\bff_{i,u}(n)$ and using \eqref{eq:QN_reform_ofdm}, the DL SINR constraint in \eqref{eq:problem_dl_ofdm} can be rewritten, and the rest of the proof is similar to Corollary~\ref{cor:DL_precoder}. 
        \qed
\end{appendices}

\bibliographystyle{IEEEtran}
\bibliography{CoMP_ADCs.bib}

\end{document}